\documentclass[10pt,a4paper]{article}
\usepackage{besrefblk}
\usepackage{color}
\usepackage{amsmath}
\usepackage{epsfig}
\usepackage{lineno}
\usepackage{multirow}
\usepackage{overpic}
\usepackage{colortbl}
\usepackage{graphicx}
\usepackage{rotating}
\usepackage{etoolbox}
\usepackage{bm}
\makeatletter
\newcommand{\wideequation}{\vspace{0cm}\preto\place@tag\kern-5cm}
\makeatother
\usepackage{float}
\usepackage[colorlinks=true]{hyperref}
\hypersetup{
linkcolor=black,}
\usepackage{geometry}
\newcommand{\psip}{\psi(2S)}

\newcommand{\jpsi}{J/\psi}
\newcommand{\elp}{e^{+}}
\newcommand{\elm}{e^{-}}

\newcommand{\mup}{\mu^{+}}
\newcommand{\mum}{\mu^{-}}

\newcommand{\prp}{p}
\newcommand{\prm}{\bar{p}}
\newcommand{\kp}{K^{+}}
\newcommand{\km}{K^{-}}

\newcommand{\Xf}{x_{f}}

\newcommand{\demoisr}{(s(1-x)-M^{2})^2+(M\Gamma)^2}

\title{Analytical formula for the cross section of hadron production from $\elp\elm$ collisions around the narrow charmouinum resonances}
\author{Ya-Nan Wang$^{1}$, Ya-Di Wang$^{1}$, Ping Wang$^{2}$ \\
   \vskip 0.5cm
{\it
 $^{1}$ North China Electric Power Unversity, Beijing 102206, China\\
 $^{2}$ Institute of High Energy Physics, Chinese Academy of Sciences, Beijing 100049, China
}
}
\begin{document}
\vskip 0.5cm

\vspace{10mm}

\maketitle



\textbf{Abstract:} 
The paper reports an analytical formula for the production cross section of $\elp\elm$ 
annihilation to hadrons in the vicinity of a narrow resonance, particularly in the $\tau$-charm 
region, while considering initial state radiation. Despite some approximations in its derivation, 
the comparison between the analytical formula and direct integration of ISR shows good accuracy, 
indicating that the analytical formula meets current experimental requirements. 
Furthermore, the paper presents a comparison of the cross section between 
the analytical formula and calculations using the {\sc ConExc} Monte Carlo generator. 
The efficiency of the analytical formula in significantly reducing computing time 
makes it a favorable choice for the regression procedure to extract the parameters 
of narrow charmonium resonances in experiments.


\section{INTRODUCTION} \label{introduction}

The parameters (such as mass $M$, total width $\Gamma$, leptonic widths $\Gamma_{ee}$ and 
$\Gamma_{\mup\mum}$, etc.) of narrow resonances, like $\jpsi$ and $\psip$, have been 
extensively discussed in theories and experiments. These resonances are often referred to 
as the "hydrogen atom" in QCD. Theoretical predications of these parameters can be made 
using different potential models or lattice QCD, 
as well as be measured from experiments with $\elp\elm$ colliders, such as BABAR, CLEO, KEDR, 
BESIII and etc.
With the availability of large datasets from experiments, the determination of these parameters 
has entered a precison era. 

\begin{figure}[htbp]
\begin{center}
    \includegraphics[angle=0,width=5.0cm, height=2.0cm]{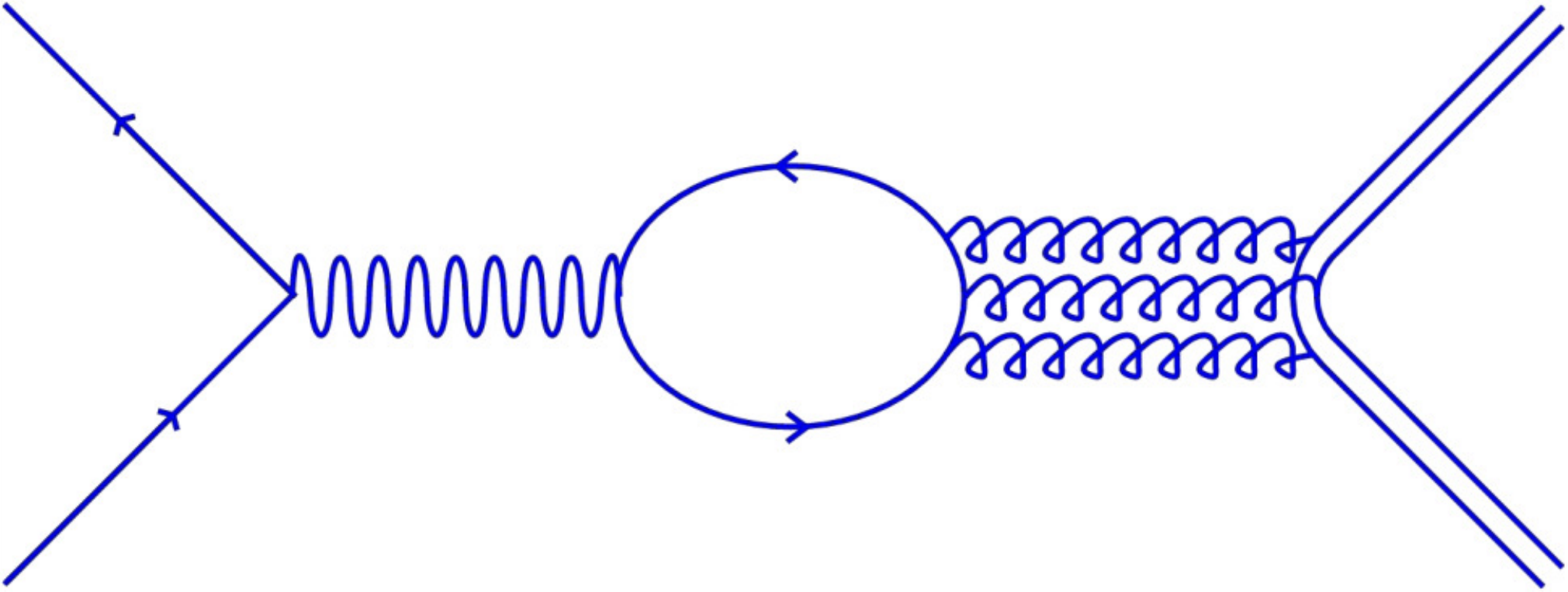}  \put(-100, 50){(a)} \put(-80,50){$A_{g}$}
    \hskip 0.5cm
    \includegraphics[angle=0,width=5.0cm, height=2.0cm]{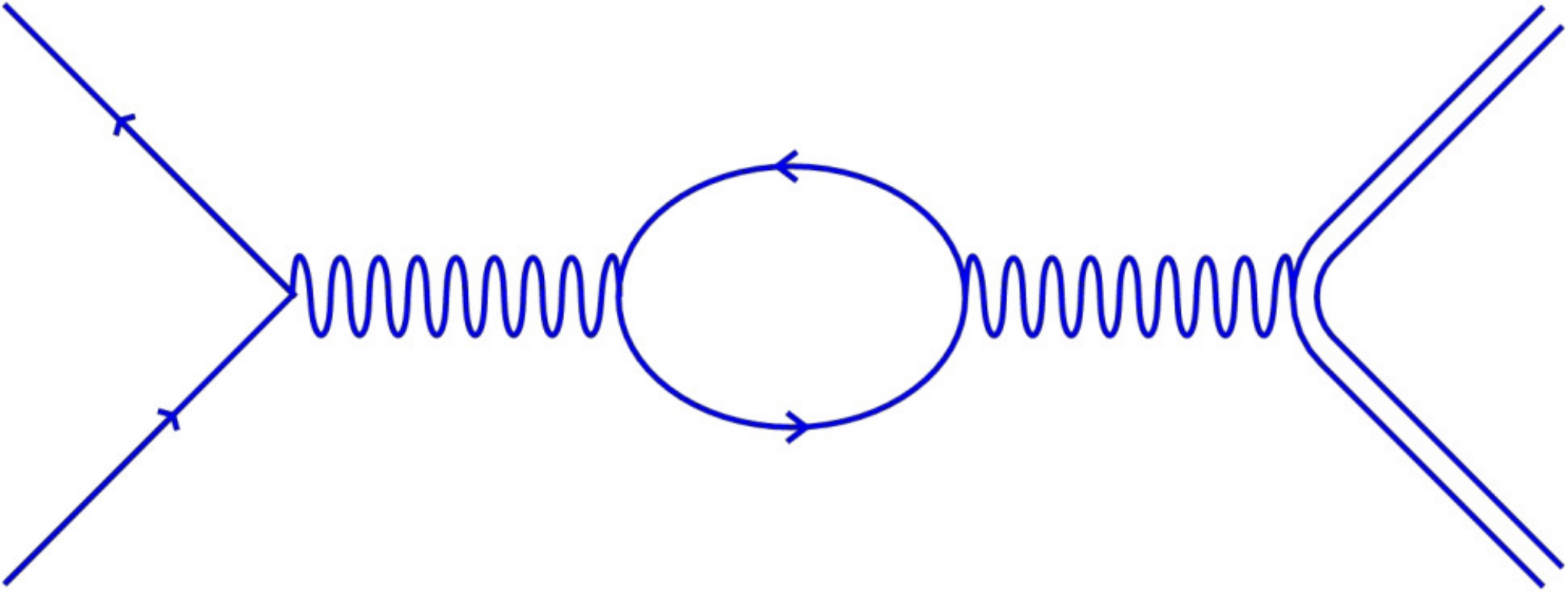}  \put(-100, 50){(b)} \put(-80,50){$A_{\gamma}$}
    \hskip 0.5cm
    \includegraphics[angle=0,width=5.0cm, height=2.0cm]{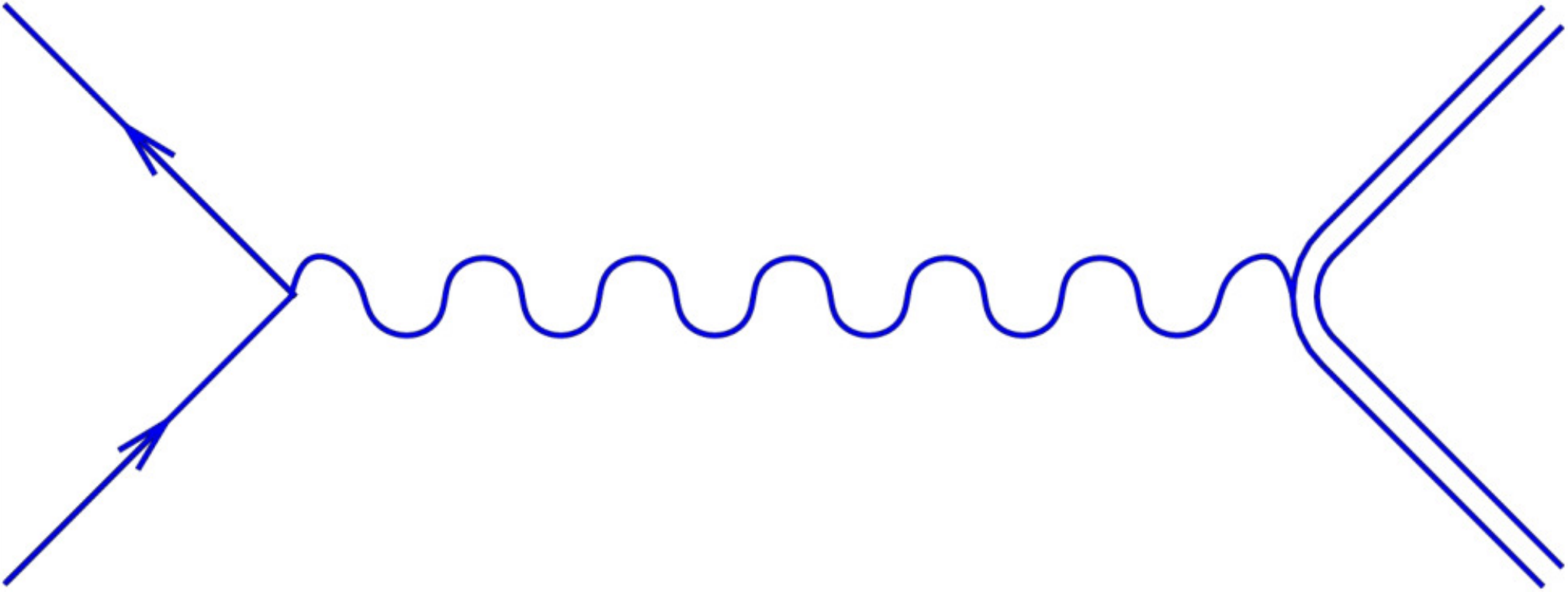}  \put(-100, 50){(c)} \put(-80,50){$A_{\rm cont}$}
 \caption{(a) $\elp\elm\to \psi(nS) \to$ hadrons via strong mechanism; (b) $\elp\elm\to\psi(nS) \to$ hadrons via EM mechanism; (c) non-resonant $\elp\elm\to$ hadrons via a vitrual photon.}
\label{fig1}
\end{center}
\end{figure}

Moreover, the branching fraction ($\mathcal{B}$) or the partial decay width ($\Gamma_{f}$) 
of a specific hadron final state plays a critical role
in understanding the mechanism of quarkonium decays and contributes to uncovering the patterns and 
properties of the quarkonium decays. 
In experimental measurements, the interference between the strong and electromagnetic amplitudes 
of $1^{--}$ resonance decays has to be considered~\cite{wangp_yuancz}.
However, it is the usual case that the interference has not been considered, or full positive or negative
interference has been subtracted in many of the past experiements. 
In fact, the interference pattern or the relative phase ($\Phi$) between strong ($A_{g}$) 
and electromagnetic ($A_{\gamma}$) amplitudes 
could be indirectly measured by comparing decay branching ratios based on SU(3) 
symmetry~\cite{jpsidecay}. The feynman diagrams for $A_{g}$ and $A_{\gamma}$ from quarkonium 
decays are depicted in Fig.~\ref{fig1} (a) and (b), respectively.
It has been conjectured that the $\Phi$ is a universal quantity, holding a constant value for 
quarkounium decays~\cite{universal}, like $\phi$, $\jpsi$, $\psip$, $\Upsilon$ and so on. 
On the contrary, the practical extraction of $\Phi$ from decays of $\phi$, $\jpsi$ and $\psip$ 
decays do not always yield consistent results~\cite{phaseresult}.
The relative phase had drawn interest from both theoretical and experimental perspectives, 
as it plays a crucial role in almost every branching ratio measurement and in the exotic 
searches~\cite{theory_interest}.
A direct measurement could be achieved by scanning experiment which introduces 
another electromagnetic amplitude from the continuum, denoted as $A_{\rm cont}$, 
as illustrated in Fig.~\ref{fig1} (c)~\cite{J5pi_phase,psipp_phase}.

Experimentally, the Born cross section for a certain hadronic decays can be expressed
as the sum of three diagrams, as depicted in Fig.~\ref{fig1}:
\begin{eqnarray}
\sigma^{0} =|A_{\rm tot}|^{2}= |A_{\rm cont}+A_{\rm \gamma}+e^{i\Phi} A_{g}|^{2}.
\end{eqnarray} 
Here, the relative phase between $A_{\gamma}$ and $A_{\rm cont.}$ is zero according to 
QED~\cite{feynman}, as confirmed by numerous experiments~\cite{KEDR,BES}.
All decay parameters ($M$, $\Gamma_{tot}$, $\Gamma_{ee}$, $\Gamma_{\mup\mum}$, $\mathcal{B}$, and 
$\Phi$) can be determined from the production cross section lineshape of $\psi(nS)$ decays
to hadronic final states. 
In the current experiemtal study, the regression procedure of parameters from the cross section 
lineshape must be accompanied with a fomula that considers the Born cross section ($\sigma^{0}$) 
convolving the initial state radiation (ISR, $F(x,S)$) and the beam energy spread 
($GS(W-W^{\prime\prime})$). The observed production cross section could be expressed as: 
\begin{eqnarray}
\label{eq_BWG}
\sigma^{\rm obs}(\sqrt{s}) = \int^{\sqrt{s}+n\Delta}_{\sqrt{s}-n\Delta} GS(\sqrt{s}-\sqrt{s^{\prime}}) d\sqrt{s^{\prime}} \int \limits_{0}^{x_f} dx F(x,s^{\prime}) \frac{\sigma^{0}(s^{\prime}(1-x))}{|1-\Pi_{0}(s^{\prime}(1-x))|^2},
\end{eqnarray}
where $\sqrt{s}$ is the energy of the center of mass system (CMS) of $\elp\elm$, 
$\Pi_{0}(s)$ represents the vacuum polarization operator~\cite{KEDR}.
The function $F(x,s)$ represents the ISR function which was first proposed 
in Ref.~\cite{Kuraev85,zlineshape}: 
\begin{eqnarray}\label{eq_isr}
F(x,s)=t x^{t-1}(1+\delta)-t(1-\frac{x}{2})+\frac{t^2}{8}\left[4(2-x)\ln\frac{1}{x}-\frac{1+3(1-x)^2}{x}\ln\frac{1}{1-x} -6+x\right],
\end{eqnarray}
where $t=\frac{2\alpha}{\pi} (\ln \frac{s}{m^{2}_{e}}-1)$, $1+\delta=1+\frac{\alpha}{\pi}(\frac{\pi^2}{3}-\frac{1}{2})+\frac{3}{4}t+t^2(\frac{9}{32}-\frac{\pi^2}{12})$, 
$x=1-\frac{s^{\prime}}{s}$, and $\sqrt{s^\prime}$ is the experimentally invariant mass of 
the final state $f$ after losing energy due to radiation. 
The upper limit $\Xf$ for $x$ corresponds to the case when $\sqrt{s^{\prime}}$ 
reaches its minimum $\sqrt{s^{min}_{f}}$, and the radiation energy reaches 
its largest value. For example, $\sqrt{s^{min}_{f}}$ could be the mass threshold of a given final state, 
or the experimental cutoff. The beam energy spread caused by the $\elp\elm$ colliders is typically
modeled as a Gaussian distribution $GS(\sqrt{s}-\sqrt{s^{\prime}})$, expressed as: 
\begin{eqnarray}\label{eq_GS}
GS(\sqrt{s}-\sqrt{s^{\prime}})=\frac{1}{\sqrt{2\pi}\Delta}e^{-\frac{(\sqrt{s}-\sqrt{s^{\prime}})^2}{2\Delta^2}},
\end{eqnarray}
where $\Delta$ represents the standard deviation of the Gaussian distribution. 
It reflects the total energy 
spread of the collider and detector, and must be determined by experiment.
For instance, in BESIII experiment~\cite{bes3},
$\Delta$ is approximately $0.9$~MeV for the $\jpsi$ resonance region, and 
about $1.3$~MeV for the $\psi(2S)$ resonance region. In a narrow energy interval, 
it could be treated as a constant.

It is evident that the cross-section calculations involve the two folds of integrations~(DTFI) 
of Eq.~(\ref{eq_BWG}). 
This can significantly slow down the regression process of experimental analysis.
In this paper, we introduce an analytical approximation formula (AAF) of the integrations of 
initial state radiation over the Born production cross section 
around a narrorw resonance in the $\tau$-charm energy region. 
It should be noted that there were some papers published 
more than 20 years ago, which presented a similar analytical formula~\cite{fengzhi,xiaohu,wangp}. 
However, the previous papers focused on the leptonic and inclusive decays, 
without considering the interference between the strong and EM amplitudes in a specific hadronic decay. 
Therefore, we present the full formula in this paper for the convenience of researchers who are 
conducting or will conduct the analysis of a single narrow resonance ~(such as $\jpsi$ or $\psip$) 
scanning at BESIII or other $\elp\elm$ collision experiments, 
especially for the measurement of the phase between strong and EM mechanisms.

The comparison results demonstrate a satisfactory level of consistency between our AAF and DTFI, 
meeting the current experimental precision requirements. However, 
it is worth noting that the computing speed has been significantly enhanced with the AAF implementation.
Throughout this paper, We use $\psi$ to represent $\jpsi$ and $\psip$ narrow resonances for brevity.

We begin with the calculation of the deduction for the AAF of the cross section 
in Section~\ref{sec2}. The comparison between the AAF and the DTFI and comparison between the AAF 
and the Monte Carlo generator are presented in Section~\ref{sec3}. 
Finally, there is a summary in Section~\ref{sec4}.

\section{Calculation of the Cross section}\label{sec2}

The Born cross section for the $\mup\mum$ final state near a resonance is:
\begin{eqnarray}
\label{eq_BW0}
\sigma^{0}(\sqrt{s})&=&\frac{4\pi\alpha^{2}}{3s}\left|1+\frac{s}{M}\frac{3\sqrt{\Gamma^{0}_{ee}\Gamma^{0}_{\mu\mu}}/\alpha}{s-M^{2}+iM\Gamma}\right|^{2}.
\end{eqnarray}
In the formula, $\Gamma^{0}_{ee}$ and $\Gamma^{0}_{\mu\mu}$ are the "bare" electronic and muonic widths.
Considering the VP effect, the dressed cross section is usually used which is part of the core of 
integrand Eq.~(\ref{eq_BWG}):
\begin{eqnarray}
\label{eq_BW}
\tilde\sigma^{0}(\sqrt{s})=\frac{\sigma^{0}(s)}{|1-\Pi_{0}(s)|^2} 
=\frac{4\pi\alpha^{2}}{3s}\frac{1}{|1-\Pi_{0}(s)|^2}\left|1+\frac{s}{M}\frac{3\sqrt{\Gamma^{0}_{ee}\Gamma^{0}_{\mu\mu}}/\alpha}{s-M^{2}+iM\Gamma}\right|^{2}.
\end{eqnarray}
Here, $\Pi_{0}(s)$ takes the similar meaning of $\Pi_0(s)$ in Ref.~\cite{KEDR}, 
which contains the contribution from lepton pairs ($\Pi_{ee}+\Pi_{\mu\mu}+\Pi_{\tau\tau}$),
hadronic decays from continuum ($\Pi_{had.}$)
and other resonances ($\Pi_{R}$) except $\psi$ resonance. 
The contribution from $\psi$ resonance production is included by the BW in Eq.~(\ref{eq_BW0}).
With definations of $\Gamma_{ee}=\Gamma^{0}_{ee}/|1-\Pi_{0}|^2$ and 
$\Gamma_{\mu\mu}=\Gamma^{0}_{\mu\mu}/|1-\Pi_{0}|^2$, 
the dressed cross section is transformed as:
\begin{eqnarray}
\label{eq_BW2}
\tilde\sigma^{0}(\sqrt{s})&=&\frac{4\pi\alpha^{2}}{3s}\left|\frac{1}{|1-\Pi_{0}(s)|}+\frac{s}{M}\frac{3\sqrt{\Gamma_{ee}\Gamma_{\mu\mu}}/\alpha}{s-M^{2}+iM\Gamma}\right|^{2}.
\end{eqnarray}
The $\Gamma_{ee}$ and $\Gamma_{\mu\mu}$ are the so called 
experimental parital widths recommended to use by the Particle Data Group~\cite{pdg}. 
In some literatures, $\Gamma_{ee}$ and $\Gamma_{\mu\mu}$ are refered to as $\Gamma^{exp}_{ee}$ 
and $\Gamma^{exp}_{\mu\mu}$.

Similarly, for a specific hadronic final state $f$, 
the dressed cross section of the hadronic production in $\elp\elm$ collision
near a resonance is:
\begin{eqnarray}
\label{eq_Born}
\tilde\sigma^{0}(\sqrt{s})&=&\beta^{2l+1}(\frac{\mathcal{F}}{s^{n/2}})^{2}\frac{4\pi\alpha^{2}}{3s}\left|\frac{1}{|1-\Pi_{0}(s)|}+(1+\mathcal{C}e^{i\Phi})\frac{s}{M} \frac{3\sqrt{\Gamma_{ee}\Gamma_{\mu\mu}}/\alpha}{s-M^{2}+iM\Gamma}\right|^{2},
\end{eqnarray}
where, the parameter $\mathcal{C}$ represents the ratio between $|A_{g}|$ and $|A_{\gamma}|$.
The dressed continuum amplitude $A_{\rm cont.}$ contributes as: 
\begin{equation}
\tilde\sigma^{\rm cont}(\sqrt{s}) = \beta^{2l+1}\frac{4\pi\alpha^{2}}{3s}(\frac{\mathcal{F}}{s^{n/2}})^{2}\frac{1}{|1-\Pi_{0}(s)|^{2}}.
\end{equation}
In the formulas, $\beta^{2l+1}$ represents the phase space factor, and
$\beta$ is the velocity of the final state particles. 
For two-body decays, 
$l$ denotes the orbital angular momentum number between 
the final state particles. For example, $l=1$ for $\kp\km$, and $l=0$ for $\prp\prm$.
The term $\mathcal{F}^{2}/s^{n}$ represents the form factor for $\elp\elm$ annihilates 
to the hadronic final state $f$. 
From a fit on the experimental data~\cite{bes3KK,babarKK,bes3ppbar,babarppbar}, 
the value of $n$ is set as 1 for meson pair and $2$ for bayron pair. 
Usually, two form factors, $G_E$ for electric and $G_M$ for magnetic, are needed to 
describe the cross section of baryon pair production. The ratio between them reaches 
unity in our region of interest, the charmonium resonance region, which is far away 
from the threshold of baryon pairs~\cite{bes3ppbar,babarppbar}. 
Thus, only one form factor is enough in this case.

Comparing with the partial width of $\psi\to \mup\mum$, the experimental partial decay 
width of $\psi\to f$ should be the value on the resonance peak ($\sqrt{s}=M$):
\begin{eqnarray}
\Gamma_{f}=\beta^{2l+1}(\frac{\mathcal{F}}{M^{n}})^{2}\Gamma_{\mu\mu}|1+\mathcal{C}e^{i\Phi}|^2
\end{eqnarray}
The branching ratio for $\psi\to f$ can be extracted with: 
\begin{eqnarray}\label{eq_br}
\mathcal{B}(\psi\to f) = \frac{\Gamma_{f}}{\Gamma_{\mu\mu}} \mathcal{B}(\psi\to \mu\mu) 
= (\frac{\mathcal{F}}{M^{n}})^{2}|1+\mathcal{C}e^{i\Phi}|^2 \frac{\Gamma_{\mu\mu}}{\Gamma}. 
\end{eqnarray} 

Considering the lepton universality in QED, $\Gamma_{ee}$ is used to substitute the term for
$\sqrt{\Gamma_{ee}\Gamma_{\mu\mu}}$. Define $A=\frac{4\pi\alpha^{2}}{3}$, 
$B=\frac{3\Gamma_{ee}}{\alpha M}$, Eq.~(\ref{eq_Born}) for hadronic production
in the absolute sign is expanded as: 
\begin{eqnarray}
\label{eq_Born2}
\tilde\sigma^{0}(\sqrt{s})
=\beta^{2l+1}(\frac{\mathcal{F}}{s^{n/2}})^{2}\left[\frac{A}{|1-\Pi_{0}(s)|^{2}s} +  
\frac{A B^2 s(1+\mathcal{C}^{2}+2\mathcal{C}\cos\Phi)}{(s-M^{2})^2+(M\Gamma)^2} + 
\frac{2AB}{|1-\Pi_{0}(s)|}\frac{(1+\mathcal{C}\cos\Phi)(s-M^2)+\mathcal{C}\sin\Phi M\Gamma}{(s-M^{2})^2+(M\Gamma)^2}\right].
\end{eqnarray}

\begin{figure}[htbp]
\begin{center}
    \includegraphics[angle=0,width=12cm, height=9cm]{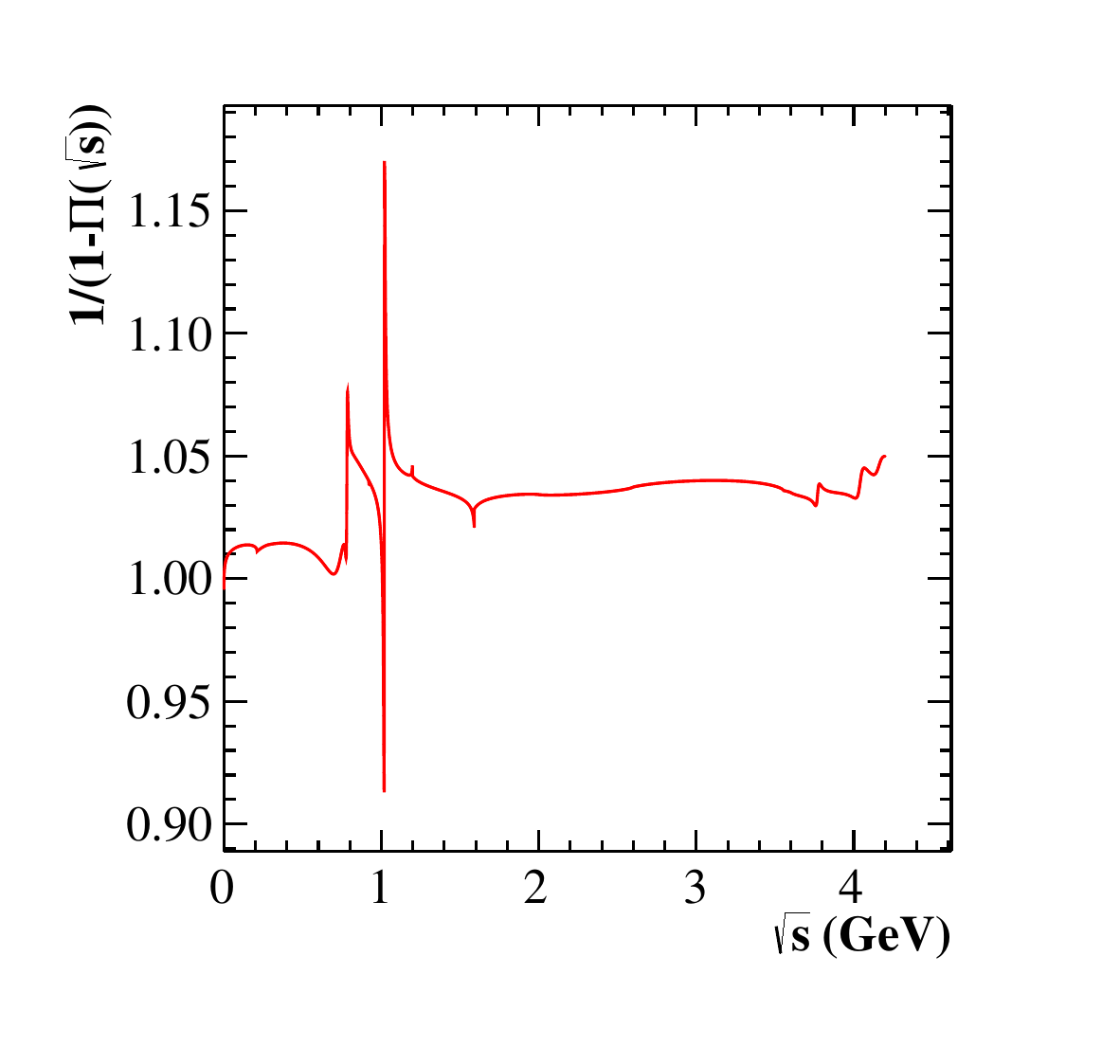}
 \caption{Distribution of $1/|1-\Pi_{0}|^{2}$ variation with $\sqrt{s}$ below $4.2$~GeV.}
\label{fig_vp}
\end{center}
\end{figure}

Considering ISR, the CMS energy squared $s$ is changed to $s(1-x)$, 
and the first integrand of Eq.~(\ref{eq_BWG}) results in the cross section after radiation:
\begin{eqnarray}
\sigma^{\rm ISR}(\sqrt{s}) &=& \int^{x_f}_0 \tilde\sigma^{\prime}(s(1-x)) F(x,s) dx 
\end{eqnarray} 
where,
\begin{eqnarray}
\sigma^{\prime}(s(1-x))=\beta^{2l+1}(\frac{\mathcal{F}}{(s(1-x))^{n/2}})^2\left[ \frac{1}{|1-\Pi_{0}(s(1-x))|^{2}} \frac{A}{s(1-x)} + \frac{C_{1}s^2-xC_{2}s^{2}}{(s(1-x)-M^{2})^2+(M\Gamma)^2} \right].
\label{eq_tot}
\end{eqnarray}
The function $F(x,s)$ has a singularity at $x=0$ due to the term $x^{t-1}$, with $t \approx 0.07$ 
at the charmonium energy region. Therefore, the integral is dominated by the value of the 
integrand with $x$ close to $0$. In the integrand, besides the BW formula, 
factors such as $1/|1-\Pi_{0}|$, phase space, and form factor are all slowly varying functions of $s$. 
Thus these factors can be approximated by their values at $x=0$.   
The $1/|1-\Pi_{0}|$ factor is calculated with a program developed in Ref.~\cite{Berends} and 
without considering the contribution of $\psi$ resonance as explained in Eq.~\ref{eq_BW}, 
and shown in Fig.~\ref{fig_vp}.
It increases the cross section by about $3\sim4\%$ at charmonium region which is 
clarified in Ref.~\cite{RCMCWG}.
The terms $C_{1}$ and $C_{2}$ are defined as: 
\begin{flalign}\label{eq_C1}
\begin{split}
C_{1}
&= \left\{ A B^2 s(1+\mathcal{C}^{2}+2\mathcal{C}\cos\Phi) + \frac{2AB}{|1-\Pi_{0}(s(1-x))|}[(1+\mathcal{C}\cos\Phi)(s-M^2)+\mathcal{C}\sin\Phi M\Gamma] \right\}/s^2   \\
&= \left\{\frac{12\pi\Gamma^{2}_{ee}}{M^2}s(1+\mathcal{C}^{2}+2\mathcal{C}\cos\Phi) + \frac{8\pi\alpha}{|1-\Pi_{0}(s(1-x))|}\frac{\Gamma_{ee}}{M}\left[(1+\mathcal{C}\cos\Phi)(s-M^2)+\mathcal{C}\sin\Phi M\Gamma\right]  \right\}/s^2 \\
&\approx \left\{\frac{12\pi\Gamma^{2}_{ee}}{M^2}s(1+\mathcal{C}^{2}+2\mathcal{C}\cos\Phi) + \frac{8\pi\alpha}{|1-\Pi_{0}(s)|}\frac{\Gamma_{ee}}{M}\left[(1+\mathcal{C}\cos\Phi)(s-M^2)+\mathcal{C}\sin\Phi M\Gamma\right]  \right\}/s^2
\end{split}
\end{flalign}

\begin{flalign}\label{eq_C2}
\begin{split}
C_{2}
&= [AB^2 s(1+\mathcal{C}^{2}+2\mathcal{C}\cos\Phi) + \frac{2ABs}{|1-\Pi_{0}(s(1-x))|} (1+\mathcal{C}\cos\Phi)]/s^{2}  \\
&= \left[ \frac{12\pi\Gamma^{2}_{ee}}{M^2}(1+\mathcal{C}^{2}+2\mathcal{C}\cos\Phi) + \frac{8\pi\alpha}{|1-\Pi_{0}(s(1-x))|}\frac{\Gamma_{ee}}{M}(1+\mathcal{C}\cos\Phi) \right]/s \\
&\approx \left[ \frac{12\pi\Gamma^{2}_{ee}}{M^2}(1+\mathcal{C}^{2}+2\mathcal{C}\cos\Phi) + \frac{8\pi\alpha}{|1-\Pi_{0}(s)|}\frac{\Gamma_{ee}}{M}(1+\mathcal{C}\cos\Phi) \right]/s
\end{split}
\end{flalign}

In Eq.~(\ref{eq_tot}), the first part is contribution from continuum cross section.
The terms with $AB^{2}$ in $C_{1}$ in Eq.~(\ref{eq_C1}) and $C_{2}$ in Eq.~(\ref{eq_C2}) 
represent for the resonace part, while the term with $2AB$ represents the interference contribution.
For our purpose, the $F(x,s)$ could be rewritten as shown in Eq.~(\ref{eq_isr2}) 
based on the approximations of 
$x^{t}=1+t \ln x$ and $\ln(1-x)=-x-x^2/2$~\cite{fengzhi}. 
With Eq.~(\ref{eq_isr2}), we can derive an analytical expression for the resonances 
and an interference with continuum contribution, 
\begin{eqnarray}\label{eq_isr2}
F(x,s)&=&x^{t-1}\cdot t(1+\delta) + x^t(-t-\frac{t^2}{4}) + x^{t+1}(\frac{t}{2}-\frac{3}{8}t^2). 
\end{eqnarray}
To proceed, some equlities must be introduced with contour integration from Ref.~\cite{anaZ}.
\begin{eqnarray}
\int^{\infty}_0 \frac{\nu x^{\nu-1}dx}{x^2+2ax\cos\theta+a^2}=a^{\nu-2}\cdot\frac{\pi\nu\sin(\theta(1-\nu))}{\sin\theta\sin\pi\nu},
\label{eq_1}
\end{eqnarray}
\begin{eqnarray}
\int^{\infty}_{\Xf} \frac{\nu x^{\nu-1}dx}{x^2+2ax\cos\theta+a^2}=\nu\left[-\frac{\Xf^{\nu-2}}{\nu-2}+2a\cos\theta\frac{\Xf^{\nu-3}}{\nu-3}-a^2(4\cos^{2}\theta-1)\frac{\Xf^{\nu-4}}{\nu-4}  \right]
(\Xf < 1).
\label{eq_2}
\end{eqnarray}
With these equlities, one can get:
\begin{eqnarray}
\int^{\Xf}_0 \frac{\nu x^{\nu-1}dx}{x^2+2ax\cos\theta+a^2}
=a^{\nu-2}\cdot\frac{\pi\nu\sin(\theta(1-\nu))}{\sin\theta\sin\pi\nu} + \nu\left[\frac{\Xf^{\nu-2}}{\nu-2}-2a\cos\theta\frac{\Xf^{\nu-3}}{\nu-3}+a^2(4\cos^{2}\theta-1)\frac{\Xf^{\nu-4}}{\nu-4} \right]. 
\label{eq_3}
\end{eqnarray}
Notice this approximation only works for $\nu<2$, so that $\frac{\infty^{\nu-2}}{\nu-2}\sim 0$.
Based on Eq.~(\ref{eq_3}), one can get:
\begin{eqnarray}
\int^{\Xf}_0 \frac{t x^{t-1}dx}{\left[s(1-x)-M^2\right]^2+M^2\Gamma^2} = \frac{1}{s^2}\cdot a^{t-2}\phi(\cos\theta,t)+\frac{t}{s^2}\cdot\left[\frac{\Xf^{t-2}}{t-2}-2a\cos\theta\frac{\Xf^{t-3}}{t-3}+a^2(4\cos^{2}\theta-1)\frac{\Xf^{t-4}}{t-4}  \right],
\label{eq_4}
\end{eqnarray}
where \begin{eqnarray}
a^2&=&(1-\frac{M^2}{s})^2 + \frac{M^2\Gamma^2}{s^2}, \notag\\
\cos\theta&=& \frac{1}{a}\cdot(\frac{M^2}{s}-1), \notag \\
\phi(\cos\theta,t)&=&\frac{\pi t\sin(\theta(1-t))}{\sin\theta\sin\pi t}. \notag
\end{eqnarray}

Based on the equlities above and ignoring the small contribution from higher orders, 
each term in the resonance and interference part in Eq.~(\ref{eq_tot}) integrated with terms 
in Eq.~(\ref{eq_isr2}) could be calculated, and the cross section considering only 
the ISR function is written as:
\begin{eqnarray}
\sigma^{\rm ISR}(\sqrt{s}) 
&=& \sigma^{\rm int1}+\sigma^{\rm int2}+\sigma^{\rm int3}+\sigma^{\rm int4}+\sigma^{\rm int5}+\sigma^{\rm cont,ISR},
\end{eqnarray} 
where each term are listed as follows, and can be calculated with the help of Eq.~(\ref{eq_4}).
In the following, $R_{2}=-2a\cos\theta$ and $R_{3}=a^{2}(4\cos^{2}\theta-1)$.
\begin{eqnarray}
\sigma^{\rm int1} &=& \beta^{2l+1}(\frac{\mathcal{F}}{s^{n/2}})^2 \int^{x_f}_0 \frac{C_{1}s^{2}}{\demoisr}  t x^{t-1} (1+\delta) dx  \notag \\
&=& \beta^{2l+1}(\frac{\mathcal{F}}{s^{n/2}})^2 C_{1}(1+\delta)\left[a^{t-2}\phi(\cos\theta,t)+t(\frac{\Xf^{t-2}}{t-2}+\frac{\Xf^{t-3}}{t-3}R_2 + \frac{\Xf^{t-4}}{t-4}R_3) \right],
\end{eqnarray}

\begin{eqnarray}
\sigma^{\rm int2} &=&\beta^{2l+1}(\frac{\mathcal{F}}{s^{n/2}})^2 \int^{x_f}_0 \frac{C_{1}s^{2}}{\demoisr}  x^t(-t-\frac{t^2}{4}) dx \notag \\ 
&=&\beta^{2l+1}(\frac{\mathcal{F}}{s^{n/2}})^2 C_{1}(-t-\frac{t^2}{4})[ \frac{a^{t-1}}{(t+1)}\phi(\cos\theta,t+1)+\frac{\Xf^{t-1}}{t-1}+\frac{\Xf^{t-2}}{t-2}R_{2} +\frac{\Xf^{t-3}}{t-3}R_{3}],
\end{eqnarray}

\begin{eqnarray}
\sigma^{\rm int3} &=&\beta^{2l+1}(\frac{\mathcal{F}}{s^{n/2}})^2 \int^{x_f}_0 \frac{-xC_{2}s^{2}}{\demoisr} t x^{t-1}(1+\delta) dx \notag \\
&=&-\beta^{2l+1}(\frac{\mathcal{F}}{s^{n/2}})^2 C_{2}t(1+\delta)\left[ \frac{a^{t-1}}{t+1}\phi(\cos\theta,t+1)+ \frac{\Xf^{t-1}}{t-1}+\frac{\Xf^{t-2}}{t-2}R_{2}+\frac{\Xf^{t-3}}{t-3}R_{3} \right],
\end{eqnarray}

\begin{eqnarray}
\sigma^{\rm int4} &=& \beta^{2l+1}(\frac{\mathcal{F}}{s^{n/2}})^2  \int^{x_f}_0 \frac{-xC_{2}s^{2}}{\demoisr}  x^{t}(-t-\frac{t^2}{4}) dx \notag \\
&\approx& \beta^{2l+1}(\frac{\mathcal{F}}{s^{n/2}})^2  C_2 (t+\frac{t^2}{4})\int^{x_f}_0\frac{x^{t+1}}{x^2+2axcos\theta +a^2} dx,
\end{eqnarray}

\begin{eqnarray}
\sigma^{\rm int5} &=&\beta^{2l+1}(\frac{\mathcal{F}}{s^{n/2}})^2  \int^{x_f}_0 \frac{C_{1}s^{2}}{\demoisr}  x^{t+1}(\frac{t}{2}-\frac{3}{8}t^2) dx \notag \\
&\approx&\beta^{2l+1}(\frac{\mathcal{F}}{s^{n/2}})^2  C_1(\frac{t}{2}-\frac{3t^2}{8})\int^{x_f}_0\frac{x^{t+1}}{x^2+2axcos\theta+a^2} dx.
\end{eqnarray}
The approximation in $\sigma^{\rm int4}$ and $\sigma^{\rm int5}$ holds water
because they only contribute no more than 1.5\% in charmonium region.
The integration in $\sigma^{\rm int4}$ and $\sigma^{\rm int5}$ could be done with simple calculus:
\begin{eqnarray}
\int^{\Xf}_0 \frac{x}{x^2+2ax\cos\theta +a^2} dx 
=-ctg\theta (tg^{-1}\frac{\Xf+a\cos\theta}{a\sin\theta}-\frac{\pi}{2} + \theta) + \frac{1}{2}\ln\frac{\Xf^2+2a\Xf\cos\theta+a^2}{a^2}.
\label{eq_5}
\end{eqnarray}
By considering the ISR effect, the QED part $\sigma^{\rm cont,ISR}$ is described as:
\begin{equation}
\sigma^{\rm cont,ISR}(\sqrt{s}) = \int^{X_{f}}_{0}F(x,s)\tilde\sigma^{cont}(s(1-x)) dx 
\end{equation}
For the contribution from continuum prosess, the beam energy spread $GS(\sqrt{s}-\sqrt{s^{\prime}})$ has 
little effect on $\sigma^{\rm cont}$ since it is almost flat in our interested region. 
Finally, the total observed cross section could be summarized as:
\begin{eqnarray}\label{eq_full2}
\sigma^{\rm obs}(\sqrt{s})&=&\int^{\sqrt{s}+n\Delta}_{\sqrt{s}-n\Delta} GS(\sqrt{s}-\sqrt{s^{\prime}})  \sigma^{\rm ISR}(\sqrt{s^\prime})d\sqrt{s^{\prime}} \notag \\
&=& \int^{\sqrt{s}+n\Delta}_{\sqrt{s}-n\Delta} GS(\sqrt{s}-\sqrt{s^{\prime}})  \left[(\sigma^{\rm int1}+\sigma^{\rm int2}+\sigma^{\rm int3}+\sigma^{\rm int4}+\sigma^{\rm int5}) \right] d\sqrt{s^{\prime}}+\sigma^{\rm cont,ISR}(\sqrt{s})
\end{eqnarray}

So far, we have derived the full AAF for the cross section of hadron production from 
$\elp\elm$ collisions around the charmouinum resonances.
In this way, the two folds of integration are reduced to one fold.
In order to verify the accuracy of the AAF, we compare the cross sections 
calculated by the AAF with those by the DTFI.
In the meanwhile, the comparison between the cross sections from AAF and those from 
Monte Carlo simulation is shown in Section~\ref{sec3}.

\section{Comparison among results from AAF, DTFI and Monte Carlo generator} \label{sec3}

\subsection{Comparison between AAF and DTFI}

Traditionally, cross sections are calculated using the DTFI, but it has been found that the DTFI 
takes a long time to calculate. For example, Table~\ref{tab_1} shows the time consumed 
if the cross section values for ten energy points be calculated, and it is about 36 seconds. 
In experimental analysis, physical parameters are extracted by a $\chi^{2}$ fit
or a likelyhood fit which means hundreds of thousands of iterations may be required.
If the initial parameters are not proper for a converged result, the regression process should be repeated.
Thus, the computing speed must be improved to meet the practical requirement.
 
The comparison of the computing time with the DTFI and the AAF is shown in Table~\ref{tab_1}.
The first column of Tab.~\ref{tab_1} is the number of energy points computed, 
and the second column is the computing time of the DTFI, while the third column of the AAF.
It is obviously that the computing time with the analytical form of the cross section is 
greatly reduced. This makes it possible that the regression process on a production cross section 
lineshape to be finished in several minutes. 
  
\begin{figure}[htbp]
\includegraphics[angle=0,width=6.5cm, height=6.5cm]{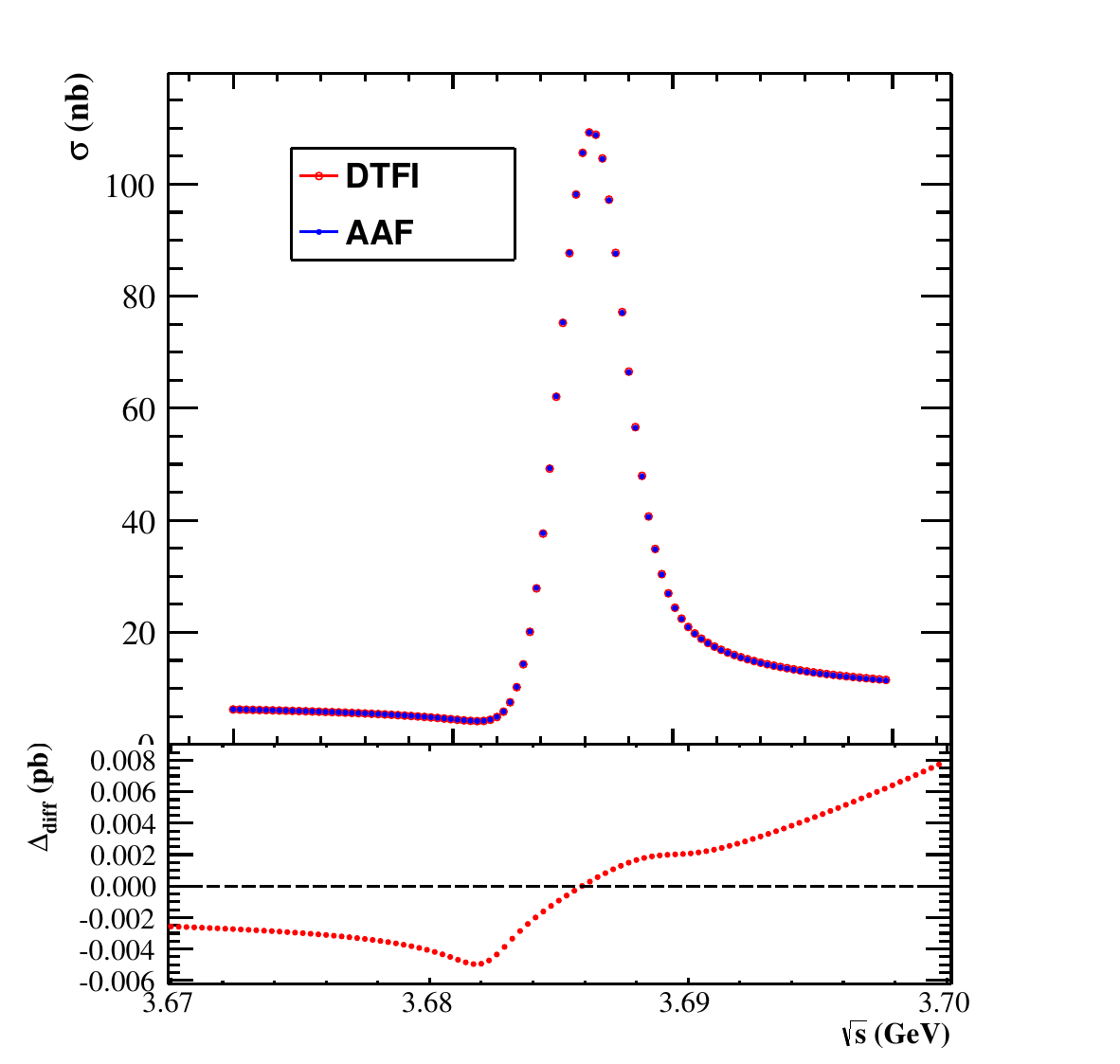}\put(-90,-5){\footnotesize $\Phi=0^\circ$}\put(-105,-5){(a)}
\includegraphics[angle=0,width=6.5cm, height=6.5cm]{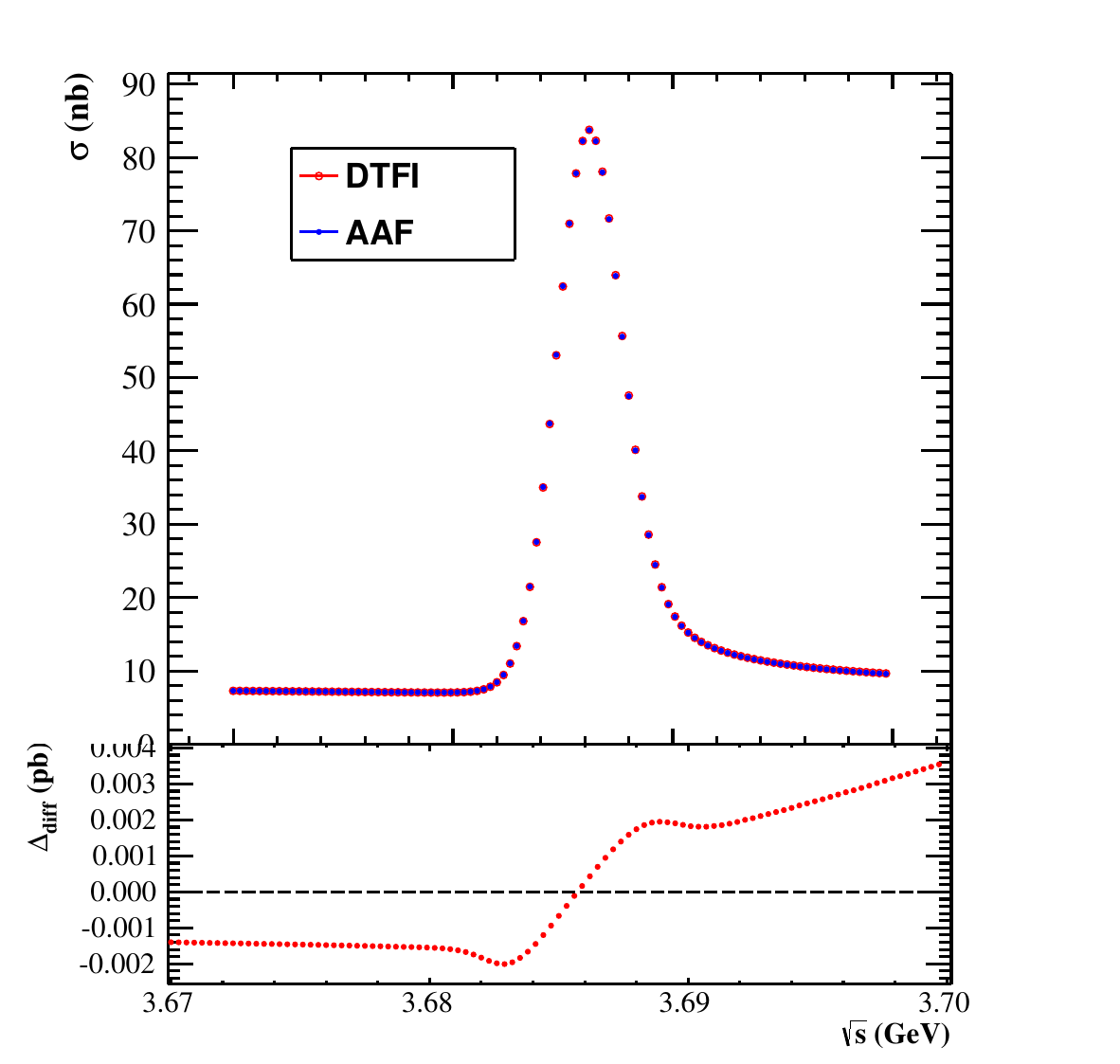}\put(-90,-5){\footnotesize $\Phi=90^\circ$}\put(-105,-5){(b)}
\includegraphics[angle=0,width=6.5cm, height=6.5cm]{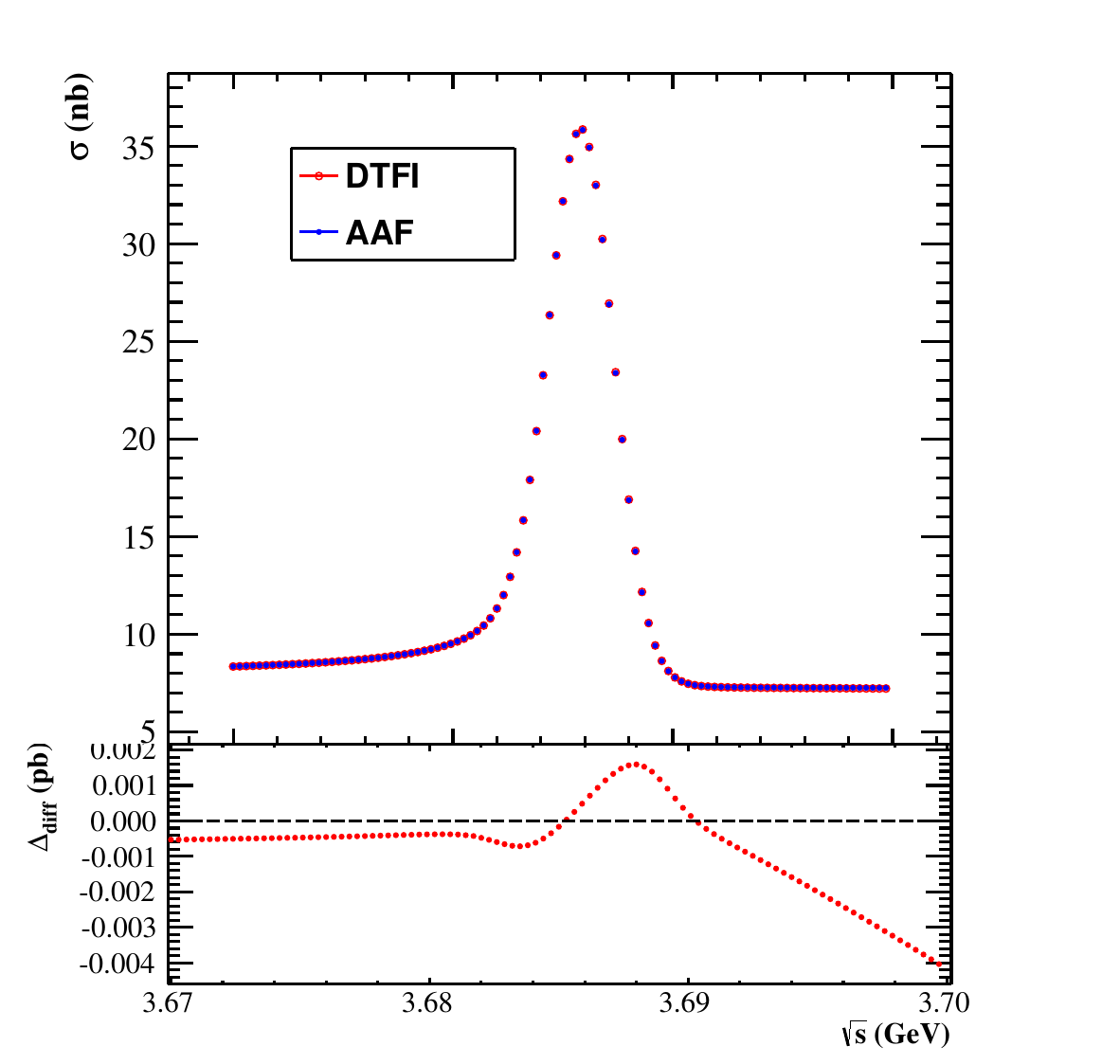}\put(-90,-5){\footnotesize $\Phi=180^\circ$}\put(-105,-5){(c)}
\caption{Comparisons of cross sections calculated with the analytic form and the DTFI in the case of
$\Phi=0^\circ$,$\Phi=90^\circ$, and $\Phi=180^\circ$, respectively.
The red dots stands for the results of the cross section calculated with the DTFI 
and blue dots the results of the cross section using the AAF. 
The bottom plots show the ratio between two results under different $\Phi$ assumptions.
}
\label{fit_result_1}
\end{figure}
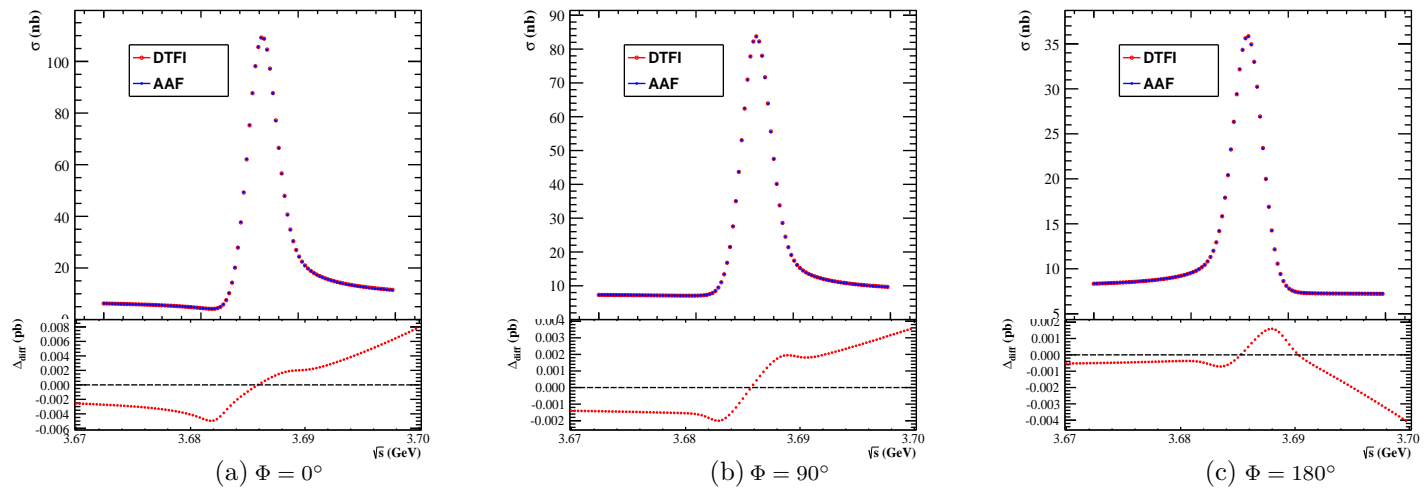

It is also necessary to compare the cross section calculated by the AAF and the DTFI to 
assess the level of precision achieved. 
Figure~\ref{fit_result_1} shows the comparison between the two forms with the parameters of 
$\psip\to\kp\km$ process as an example.
The top figures in (a), (b) and (c) represent the results under different asumptions of 
$\Phi$: (a) $\Phi=0^\circ$, (b) $\Phi=90^\circ$, (c) $\Phi=180^\circ$. 
The red and blue dots represent for the results of 
the cross section calculated using the DTFI and the AAF, respectively. 
The difference is calculated as 
$\delta=\frac{(\sigma^{\rm DTFI}-\sigma^{\rm AAF})}{(\sigma^{\rm DTFI}+\sigma^{\rm AAF})/2.0}$ and 
presented at the bottom of each figure.
Here, $\sigma^{\rm DTFI}$ stands for cross section calculated using the DTFI, while 
$\sigma^{\rm AAF}$ denotes cross section calculated using the AAF.
From comparison, the difference between the two forms is less than 1\% under different 
assumptions of $\Phi$. 

\begin{table}[htp]
\begin{center}
\caption{The comparison of the computing time with the DTFI and the AAF are listed in this table. 
The first column is the number of energy points computed $N_{points}$,
the second column the consuming time of the DTFI ($T_{\rm DTFI}$), and the third column 
of the AAF ($T_{\rm AAF}$).} \label{tab_1}
\begin{tabular}{|c|c|c|c|c|c|c|} \hline
$N_{points}$    &$T_{\rm DTFI}$~(second)     &$T_{\rm AAF}$~(second)\\ \hline
$10$            &35.9                    &0.3\\ \hline
$100$           &385.3                   &0.7 \\ \hline
$500$           &1721.4                  &2.0\\ \hline
$1000$          &3988.5                  &3.9\\ \hline
\end{tabular}
\end{center}
\end{table}

\begin{figure}[htbp]
\includegraphics[angle=0,width=6.5cm, height=5.5cm]{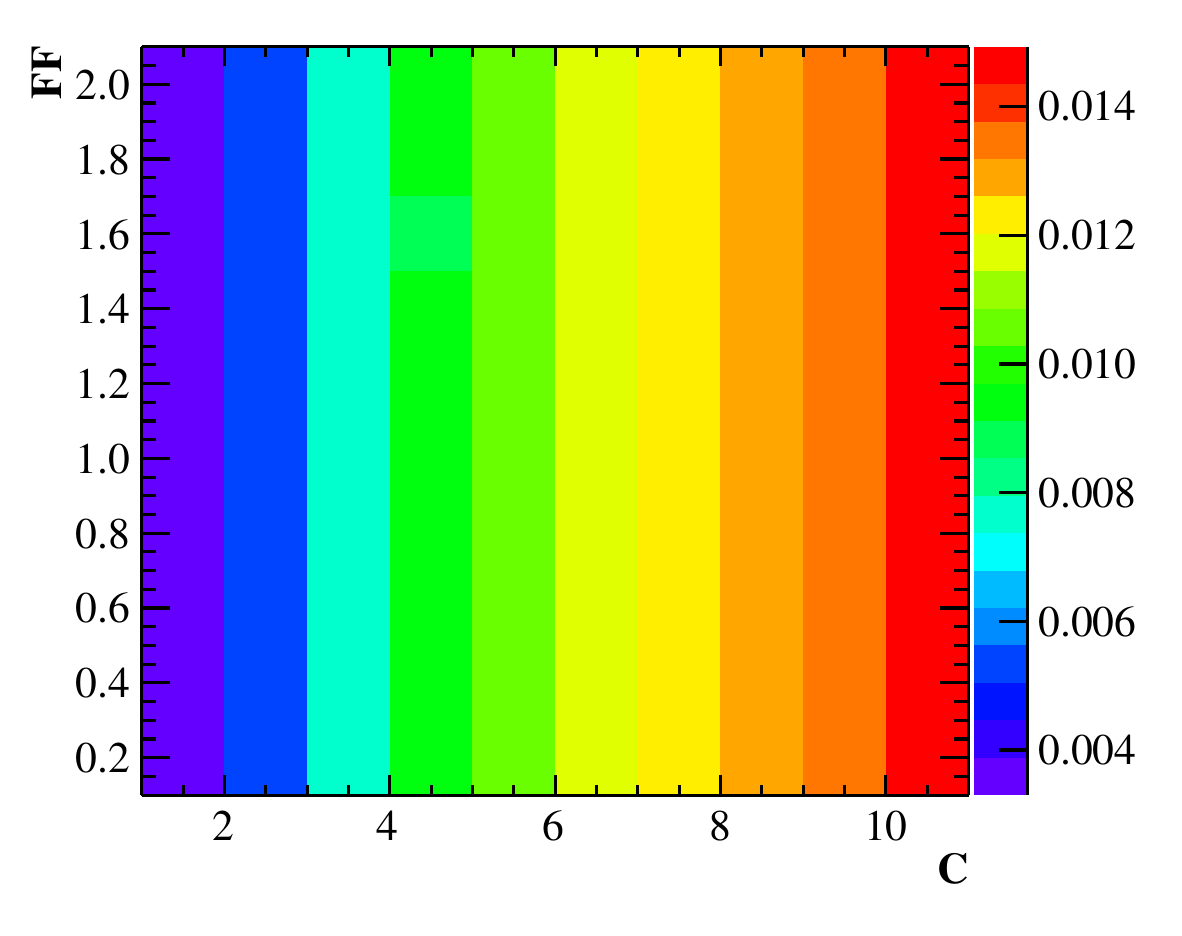}\put(-90,-5){\footnotesize $\Phi=0^\circ$}\put(-105,-5){(a)}
\includegraphics[angle=0,width=6.5cm, height=5.5cm]{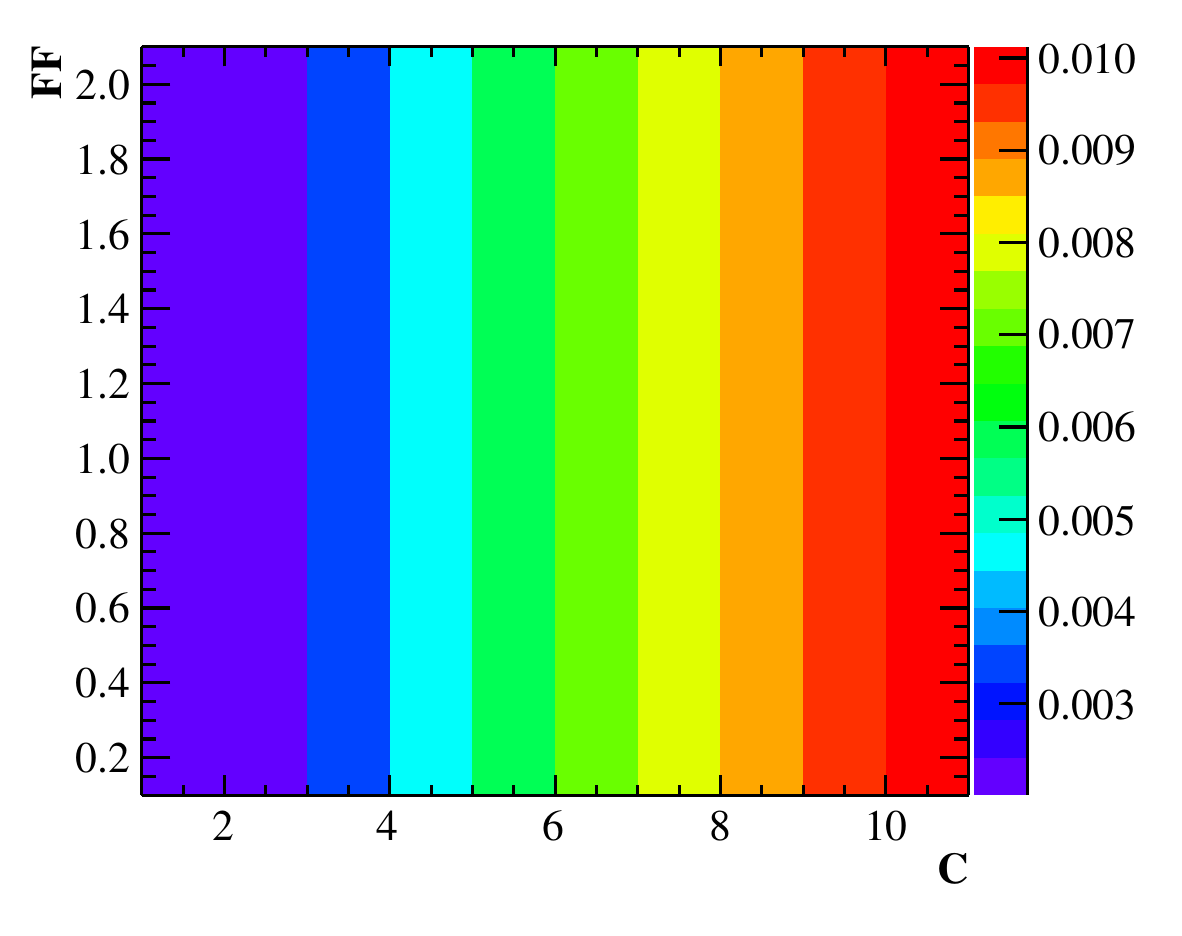}\put(-90,-5){\footnotesize $\Phi=90^\circ$}\put(-105,-5){(b)}
\includegraphics[angle=0,width=6.5cm, height=5.5cm]{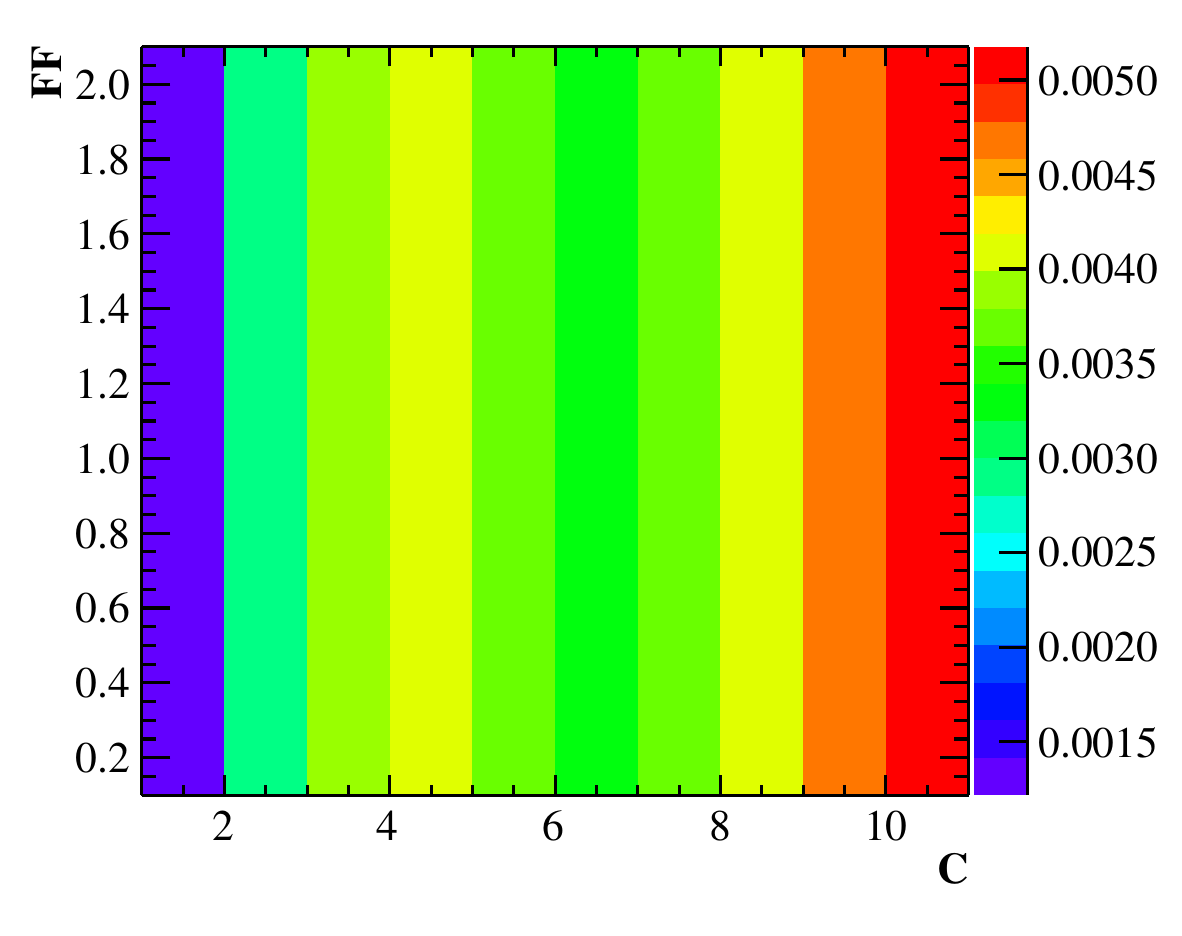}\put(-90,-5){\footnotesize $\Phi=180^\circ$}\put(-105,-5){(c)}
\caption{Difference between DTFI and AAF with $n=1$ in different assumptions of 
$\mathcal{C}$, $\mathcal{F}$ and $\Phi$.
The color in each box stands for the largest differnce between $3.67$~GeV to $3.70$~GeV. 
}
\label{fig_cpr_meson}
\end{figure}
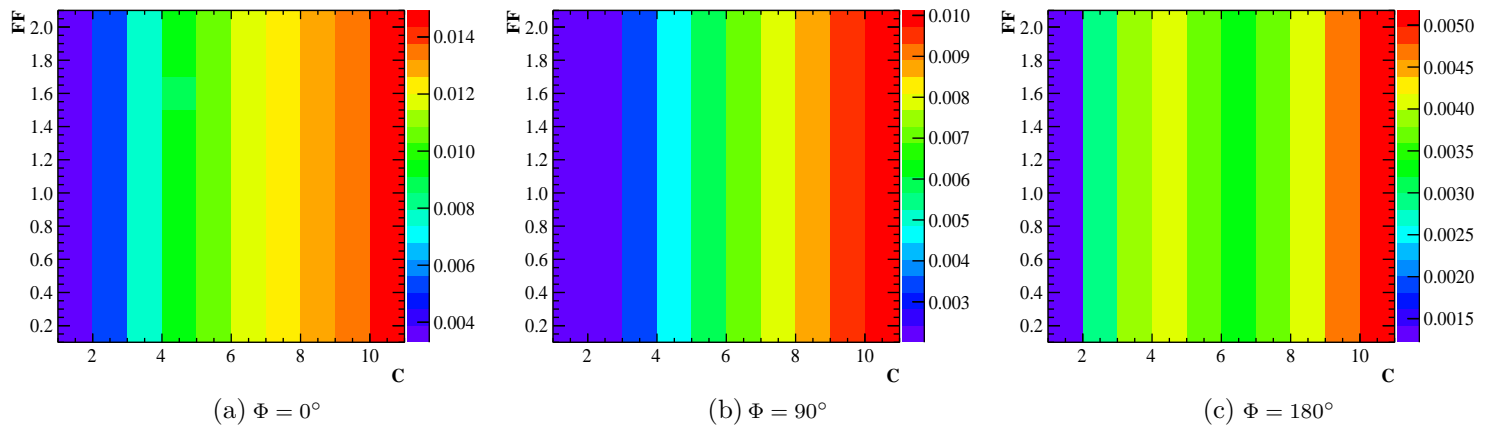

\begin{figure}[htbp]
\includegraphics[angle=0,width=6.5cm, height=5.5cm]{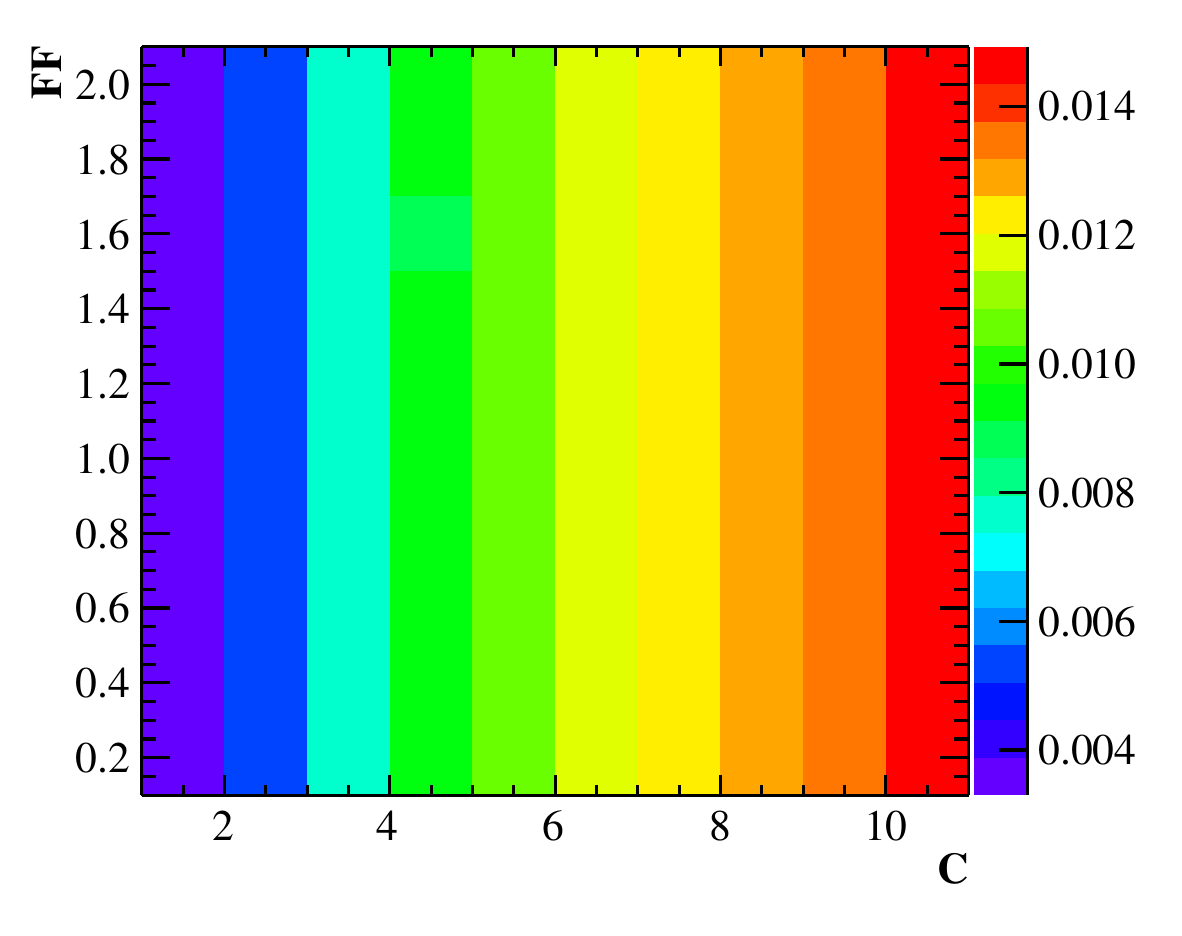}\put(-90,-5){\footnotesize $\Phi=0^\circ$}\put(-105,-5){(a)}
\includegraphics[angle=0,width=6.5cm, height=5.5cm]{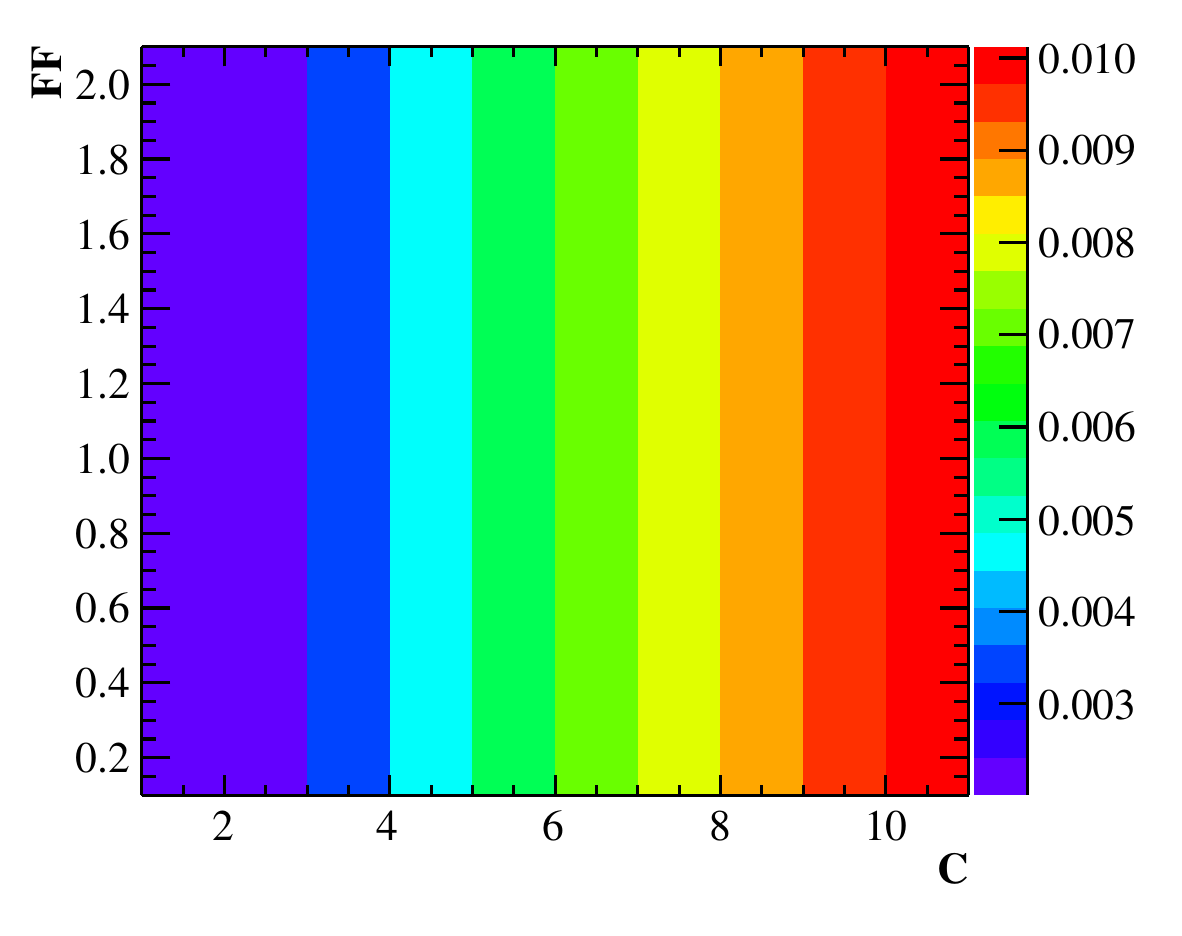}\put(-90,-5){\footnotesize $\Phi=90^\circ$}\put(-105,-5){(b)}
\includegraphics[angle=0,width=6.5cm, height=5.5cm]{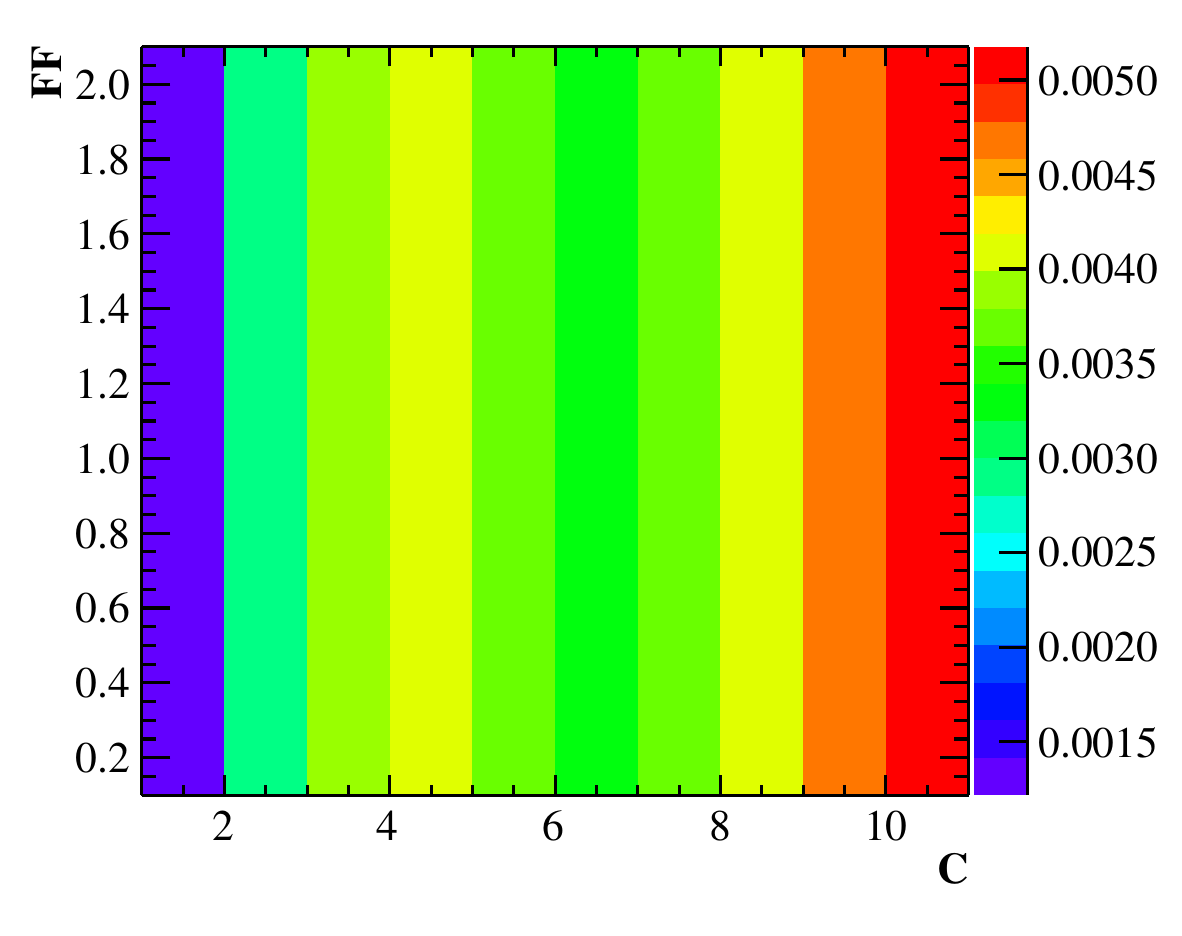}\put(-90,-5){\footnotesize $\Phi=180^\circ$}\put(-105,-5){(c)}
\caption{Difference between DTFI and AAF with $n=2$ in different assumptions of 
$\mathcal{C}$, $\mathcal{F}$ and $\Phi$.
The color in each box stands for the largest differnce between $3.67$~GeV to $3.70$~GeV. 
}
\label{fig_cpr_bb}
\end{figure}

To gain a deeper understanding of the differences between DTFI and AAF, 
a careful comparison with $n=1$ for a meson pair is implemented. 
The $\mathcal{C}$ and $\mathcal{F}$ parameters are changed under 
different assumptions of $\Phi$. The largest difference between DTFI and AAF in range of 
$(3.67,3.70)$~GeV is highlighted in color and shown in Fig.~\ref{fig_cpr_meson}.
The comparison with $n=2$ for a bayron pair is shown in Fig.~\ref{fig_cpr_bb}. 
From the comparison, we observe that the difference for $n=1$ is no more than 1.5\%, while 5\% for $n=2$. 
This is probably caused by the different position of form factor in the DTFI and AAF. In DTFI,
the form factor is merged directly in the Born cross section, while in AAF, the form factor is 
multiplied after the ISR and Gauss resolution integrations. 
From Fig.~\ref{fig_cpr_meson} and Fig.~\ref{fig_cpr_bb}, it is evident that the influence of 
$\mathcal{C}$ is much larger than that of $\mathcal{F}$. 
This maybe due to the fact that the cross section changes dramatically around the resonance
according to $\mathcal{C}$, while changes relatively slowly according to $\mathcal{F}$.

There might be a concern about the influence from the difference between the DTFI and AAF for the 
parameters determination in experiments. An input-output (IO) check is implemented with DTFI 
cross section values at nine energy points around $\psi(2S)$ which decays to $\kp\km$ and $\prp\prm$
final states. The regression procedure is finished with AAF. 
The total uncertainty is set at 5\% for all energy points, which is less than or roughly equal to
the current practical experimental data around resonance. The output values
for each parameters are listed in Table~\ref{tab_2}. From the IO check, no bias is observed 
within the uncertainty.
\begin{table}[htp]
\centering
\caption{Input-output check results with $\kp\km$ and $\prp\prm$ pair parameters 
under different assumptions of $\Phi$.}
\label{tab_2}
\begin{tabular}{|c|c|c|c|c|c|c|c|c|c|c|c|} \hline
\multirow{2}{*}{$\kp\km$ Parameter}      &\multirow{2}{*}{Input}  &\multicolumn{3}{c|}{Output}  \\ \cline{3-5}
                  &        &$\Phi=0^{\circ}$     &$\Phi=90^{\circ}$    &$\Phi=180^{\circ}$  \\ \hline
$\mathcal{C}$  &$3.0$      &$2.96\pm0.32$        &$2.94\pm0.21$        &$3.01\pm0.81$  \\ \hline
$\mathcal{F}$  &$0.5$      &$0.509\pm0.008$      &$0.509\pm0.009$      &$0.507\pm0.008$  \\ \hline
$\Phi$         &$-$        &$-5.9\pm55.2$        &$89.7\pm7.8$         &$185.3\pm126.7$  \\ \hline
$\chi^2/ndf$   &$-$        &$0.08/6$             &$0.00/6$             &$0.02/6$   \\  \hline
\multirow{2}{*}{$\prp\prm$ Parameter}      &\multirow{2}{*}{Input}  &\multicolumn{3}{c|}{Output}  \\ \cline{3-5}
                  &        &$\Phi=0^{\circ}$     &$\Phi=90^{\circ}$    &$\Phi=180^{\circ}$  \\ \hline
$\mathcal{C}$  &$15$       &$15.12\pm0.35$       &$14.97\pm0.59$       &$14.89\pm0.78$  \\ \hline
$\mathcal{F}$  &$2.5$      &$2.49\pm0.04$        &$2.51\pm0.07$        &$2.51\pm0.07$  \\ \hline
$\Phi$         &$-$        &$-1.2\pm25.7$        &$88.5\pm6.2$         &$169.2\pm38.6$  \\ \hline
$\chi^2/ndf$   &$-$        &$0.67/6$             &$0.32/6$             &$0.08/6$   \\  \hline
\end{tabular}
\end{table}

\subsection{Comparison between AAF and {\sc ConExc} Monte Carlo generator}

Since the experimental analysis is performed with regression and simulation, 
it is very necessary to check the consistency between them.
Otherwise, the parameters from the analysis will never converge no matter 
how many iterations one does. At BESIII, the {\sc ConExc} generator is developed to take into account 
the ISR effect up to the next-leading-order~\cite{conexc}.
It is widely used with at least seventy hadronic decay modes implemented 
with effective center-of-mass energy coverage from the two pion mass threshold up to about 6 GeV.
The accuracy achieved for the ISR correction reaches the level achieved by the 
KKMC generator~\cite{kkmc}.
Furthermore, the {\sc ConExc} is used for the R-value and light meson resonance measurements at BESIII. 
The cross section with ISR calculated by the {\sc ConExc} Monte Carlo generator is written as:
 \begin{eqnarray}
 \sigma(\sqrt{s})\equiv \sigma^{\mathrm{I}}(\sqrt{s})+\sigma^{\mathrm{II}}(\sqrt{s}),
 \end{eqnarray}
where
 \begin{eqnarray}
 \sigma^{\mathrm{I}}(\sqrt{s}) = \int^{M_0}_{M_{th}} dm \frac{2m}{s}W(x,s)\sigma^0(\sqrt{s}) \label{eq66}
 \end{eqnarray}
 \begin{eqnarray}
 \sigma^{\mathrm{II}}(\sqrt{s}) = \int^{\sqrt{s(1-b)}}_{M_0} dm \frac{2m}{s}W(x,s)\sigma^0(\sqrt{s}) +\sigma^0(\sqrt{s})\lim_{a \to 0}\int^{a}_{b} W(x,s) dx  \label{eq67}
 \end{eqnarray}
The mass threshold for hadronic final states is $M_{th}$, and the integral is composed of 
two parts which are separated by a point $M_0$=$\sqrt{s-2\sqrt{s}E^{cut}_{\gamma}}$ with 
an energy cut $E^{cut}_{\gamma}$ on the ISR photon. And $b$ is a small value, $b\ll 1$. 
In practice, $E^{cut}_{\gamma}$ is set to the energy sensitivity of photon detection. 
Here, since the {\sc ConExc} only consider the ISR effect, the $\sigma^0$ must be the 
cross section after convolving the Gaussian resolution to consider both beam energy spread 
and ISR effects. The radiative function W$(x,s)$ is expressed as:
 \begin{eqnarray}
W(x,s)=\Delta \beta x^{\beta -1}-\frac{\beta}{2}(2-x)+\frac{\beta ^2}{8}\left\{(2-x)[3\ln(1-x)-4\ln x]-4\frac{\ln(1-x)}{x}-6+x\right\},
 \end{eqnarray}
which has some difference with Eq.~(\ref{eq_isr}) or Eq.~(\ref{eq_isr2}). 
The $\beta$ takes the same defination as $t$ in Eq.~(\ref{eq_isr}) or Eq.~(\ref{eq_isr2}).
The $\Delta$ is in a different representation from $1+\delta$ in Eq.(\ref{eq_isr}) 
or Eq.~(\ref{eq_isr2}).
The order of convolutions of ISR and beam energy spread in the AAF is exactly the opposite to 
that in the {\sc ConExc} generator. 
In Ref.~\cite{convolution_order}, it has been pointed out that
the order of convolutions does not affect the results in a narrow energy range and with a small $\Xf$.
A test is performed with different orders around $\psi(2S)$ resonance with $\Xf=0.1$. The difference
between two orders is no more than 0.05\% with the same parameterizations as that used for 
Fig.~\ref{fig_cpr_meson} and Fig.~\ref{fig_cpr_bb}.

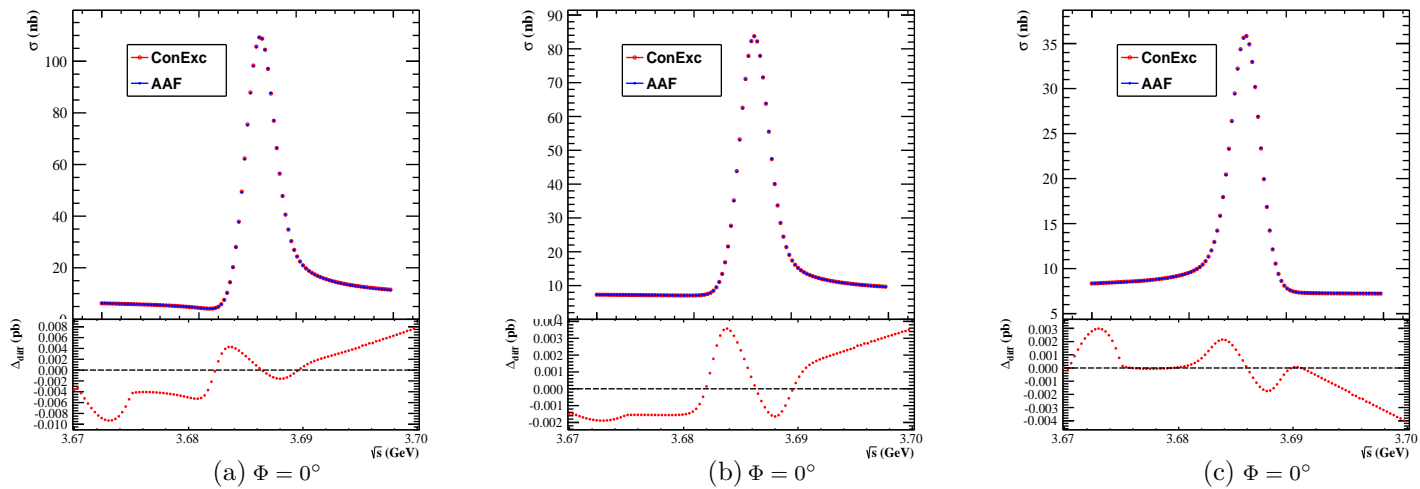
\begin{figure}[htbp]
\includegraphics[angle=0,width=6.5cm, height=6.5cm]{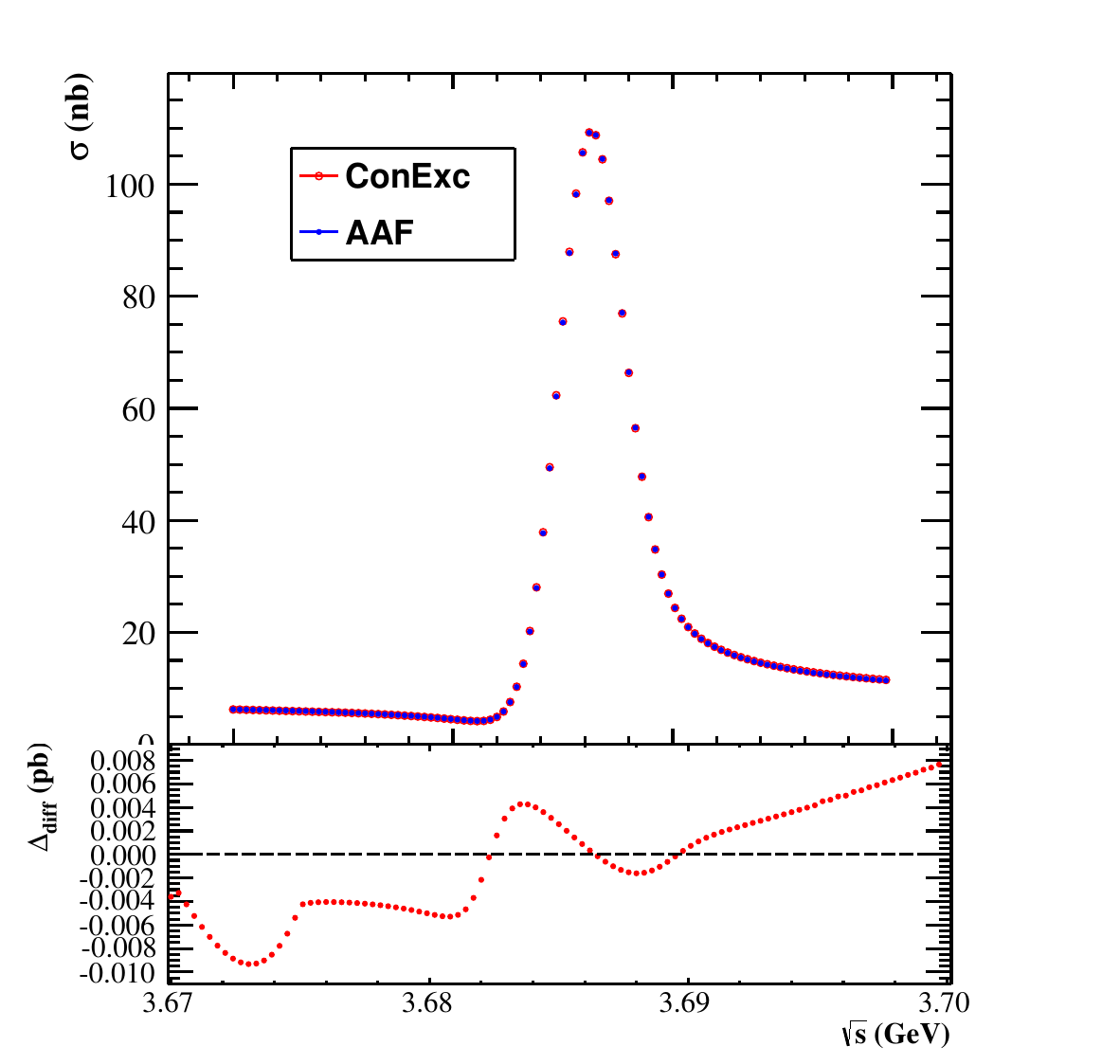}\put(-90,-5){\footnotesize $\Phi=0^\circ$}\put(-105,-5){(a)}
\includegraphics[angle=0,width=6.5cm, height=6.5cm]{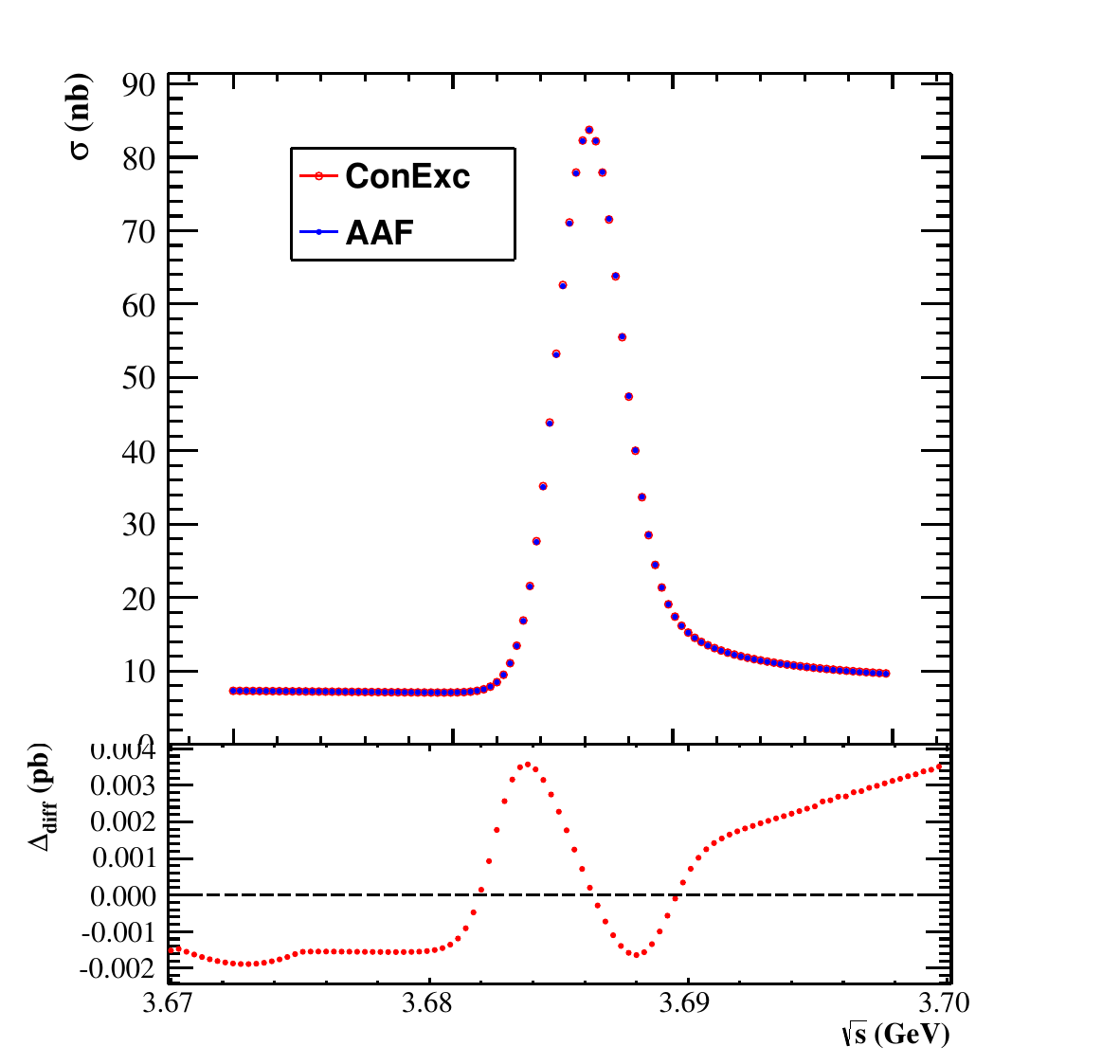}\put(-90,-5){\footnotesize $\Phi=0^\circ$}\put(-105,-5){(b)}
\includegraphics[angle=0,width=6.5cm, height=6.5cm]{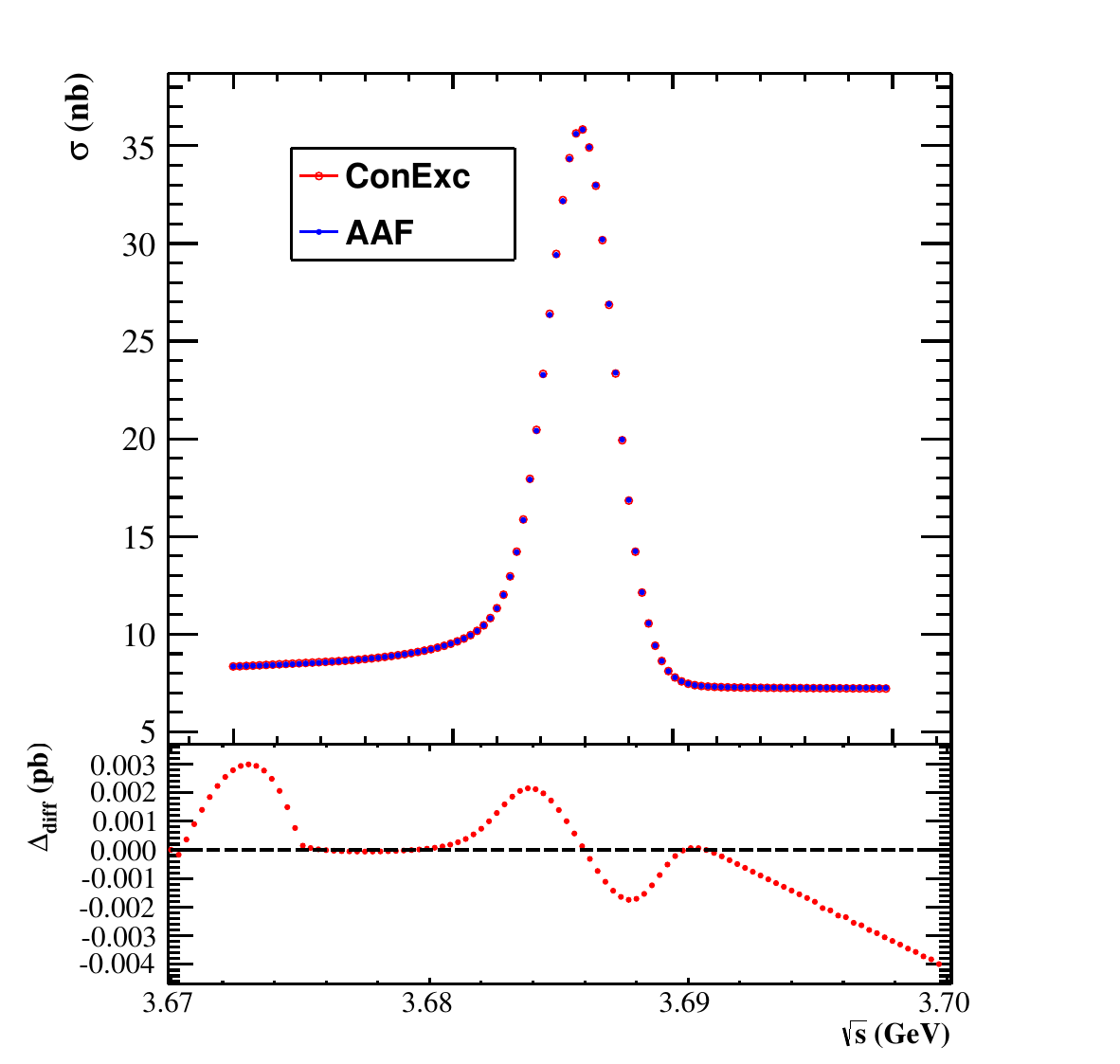}\put(-90,-5){\footnotesize $\Phi=0^\circ$}\put(-105,-5){(c)}
\caption{Comparisons of the analytic form and {\sc ConExc} generator in the case of $\Phi=0^\circ$,$\Phi=90^\circ$, and $\Phi=180^\circ$, respectively.
The red dots stand for the results of the cross section calculated with the {\sc ConExc} generator 
and blue dots the results of the cross section by using the analytic form.
The bottom plots show the ratio between two results under different $\Phi$ assumptions.
}
 \label{fit_result_2}
\end{figure}

In contrast to the AAF, the {\sc ConExc} generator computes the cross section using a sampling method.
Additionally, the radiative function $W(x,s)$ in {\sc ConExc} generator is slightly different 
from $F(x,s)$ in the AAF.
It is necessary to compare the analytic form with the {\sc ConExc} Monte Carlo generator under 
different assumptions of $\Phi$.
Figure~\ref{fit_result_2} shows the results of the cross section calculated using the {\sc ConExc} 
generator and the AAF in the cases of $\Phi=0^\circ$,$\Phi=90^\circ$, and $\Phi=180^\circ$
for the $\elp\elm\to\kp\km$ channel. 
The red dots and blue dots represent the results of
the cross section calculated by using the {\sc ConExc} generator and the AAF, respectively.
Moverover, the ratio of two cross sections is calculated as 
$ R=\frac{\sigma^{\rm\sc ConExc}}{\sigma^{\rm AAF}}$ and presented in Fig.~\ref{fit_result_2}.
Here, $\sigma^{\rm\sc ConExc}$ stands for the cross section calculated by using the {\sc ConExc} 
Monte Carlo generator, while $\sigma^{\rm AAF}$ denotes the cross section calculated by using
the AAF. The comparison result shows that the difference between the two cross section values is 
no more than 1\% under different assumptions of $\Phi$ for the $\kp\km$ pair.

\section{Conclusions}\label{sec4}

In summary, the analytical formula for the cross section of hadron production from 
$\elp\elm$ collisions 
around the narrow charmouinum resonances in $\tau$-charm energy region has been presented.
Comparisons between the analytical formula and the direct two folds of integrations shows 
good accuracy, with the difference between the two cross section results for meson pairs 
being no more than 1.5\%, and no more than 5\% for bayron pairs. 
Based on the IO check,  the observed difference will not introduce any bias in parameter 
determination for experiments involving cross sections with a 5\% uncertainty. 
Users are advised to conduct the IO check independently for the specific physical channel
in question. 
Additionally, the comparison in the cross section between the analytical formula and 
the {\sc ConExc} generator also shows good consistency. 
Most importantly, the analytical formula can greatly shorten
the computing time, which is of great significance for experiments to extract the parameters of
narrow charmonium resonances, and helpful for efficient determinations 
for each branching 
ratios. Finally, for those interested in the AAF, more information can be found at the link: 
https://github.com/yakuma320/phase\_measurement/blob/main/AAF.C.

{\it The authors would like to express their gratitude to the staff of BESIII and 
the IHEP computing center for their strong support. 
We are grateful to Prof. Changzhang Yuan for his suggestion as well as Prof. Xiaohu Mo 
for their kind help and benificial discussions.}



\newpage

\end{document}